\documentclass[aip,jap,reprint,twocolumn,nobalancelastpage]{revtex4-2} 
\usepackage{graphicx}
\usepackage{amsmath}
\usepackage{amsfonts}
\usepackage{txfonts}%\usepackage[upint]{newtx}
\usepackage{upgreek}
\usepackage{circuitikz} % needed for figure 1

\makeatletter

\pgfcircdeclarebipole{}
    {\ctikzvalof{bipoles/tline/height}}{tline}
    {\ctikzvalof{bipoles/tline/height}}
    {\ctikzvalof{bipoles/tline/width}}
{   
    \pgfpointdiff{\pgfpointanchor{\ctikzvalof{bipole/name}start}{center}}
                 {\pgfpointanchor{\ctikzvalof{bipole/name}end}{center}}
    \pgfmathparse{veclen(\the\pgf@x,\the\pgf@y)}
    \pgftransformxshift{\pgfmathresult/2-32pt}
    \pgf@circ@res@left=\dimexpr-\pgfmathresult pt+50pt\relax

    \pgf@circ@res@step=.2\pgf@circ@res@right % half x axis
    \pgfsetlinewidth{\pgfkeysvalueof{/tikz/circuitikz/bipoles/thickness}\pgfstartlinewidth}
    \pgfpathmoveto{\pgfpoint{\pgf@circ@res@right-\pgf@circ@res@step}{\pgf@circ@res@up}}
    \pgfpathlineto{\pgfpoint{\pgf@circ@res@left+\pgf@circ@res@step}{\pgf@circ@res@up}}
    \pgfpatharc{-90}{90}{-\pgf@circ@res@step and -\pgf@circ@res@up}
    \pgfpathlineto{\pgfpoint{\pgf@circ@res@right-\pgf@circ@res@step}{\pgf@circ@res@down}}
    \pgfsetfillcolor{white}
    \pgfusepath{draw,fill}

    \pgfpathellipse{\pgfpoint{\pgf@circ@res@right-\pgf@circ@res@step}{0}}
                   {\pgfpoint{\pgf@circ@res@step}{0}}
                   {\pgfpoint{0}{-\pgf@circ@res@up}}
    \pgfsetfillcolor{white}
    \pgfusepath{draw, fill}

    \pgfsetlinewidth{\pgfstartlinewidth}
    \pgfpathmoveto{\pgfpoint{\pgf@circ@res@right-1*\pgf@circ@res@step}{0pt}}
    \pgfpathlineto{\pgfpoint{\pgf@circ@res@right}{0pt}}
    \pgfusepath{draw}
}
\def\centerarc[#1](#2)(#3:#4:#5)% [draw options] (center) (initial angle:final angle:radius) 
{ \draw[#1] ($(#2)+({#5*cos(#3)},{#5*sin(#3)})$) arc (#3:#4:#5); }
\usepackage{hyperref}
\hypersetup{colorlinks,linkcolor=blue,citecolor=blue,urlcolor=blue}
\newcommand{\vect}[1]{\mathbf#1}
\newcommand{\sigc}{\sigma_c}
\newcommand{\vects}[1]{\boldsymbol{#1}}
\newcommand{\arccosh}{\cosh^{-1}}
\newcommand{\arctanh}{\tanh^{-1}}
\renewcommand{\Re}{\mathrm{Re}}
\renewcommand{\Im}{\mathrm{Im}}
\renewcommand{\arctan}{\tan^{-1}}

\begin{document}
%------------------------------------------------------------------------------

\title{DC Power Transported by Two Infinite Parallel Wires}
\author{\rule{0pt}{12pt}Marc Boul\'e}
\affiliation{\'Ecole de technologie sup\'erieure, Montr\'eal, Qu\'ebec, Canada H3C~1K3}
\date{\today}

\begin{abstract}
\begin{minipage}{5.5in}
This paper presents the calculation of the electrical power transported by the electromagnetic fields of two parallel wires carrying opposite DC currents. The Poynting vector is developed in bipolar coordinates and symbolically integrated over different surfaces. For perfectly conducting wires, the purely longitudinal power in the space surrounding the wires is shown to be equal to that which is produced by the battery (and consumed by the load resistor). For resistive wires, the longitudinal power transported by the fields is shown to diminish according to the distance traveled, and the loss is proved to be equal to the power entering the wires via the fields at their surfaces.\\
\textit{\small This article may be downloaded for personal use only. Any other use requires prior permission of the author and AIP Publishing. This article appeared in {\normalfont American Journal of Physics \textbf{92}(1), pp. 14-22 (2024)}, and may be found at} {\footnotesize\url{https://doi.org/10.1119/5.0121399}}.
\end{minipage}
\end{abstract}

\maketitle

%------------------------------------------------------------------------------
\section{Introduction}
\label{sec_intro}
%------------------------------------------------------------------------------

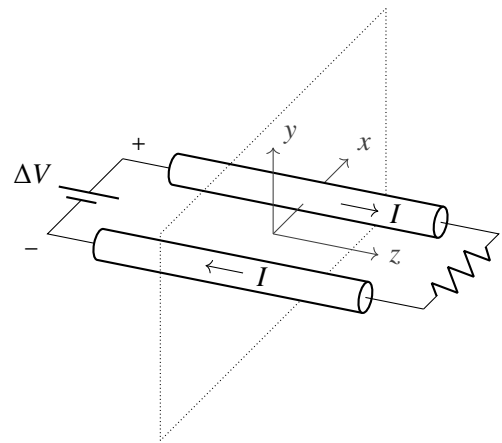
\begin{figure}[b]
\centerline{\begin{circuitikz}[scale=1,smallR/.style={R, resistors/scale=0.8}]
\def\scl{1.2}
% x axis
\draw[darkgray, ->] (0,0) -- (1,1) node[anchor=south west] {\scalebox{\scl}{$x$}};
% plane back
\draw[densely dotted] (1.5,3) -- (1.5,0.25) -- (-1.5,-2.75);
% top line and resistor
\draw (-2,1) 
to [tline] (3,0)
to [smallR] (2, -1);
% battery and bottom line
\draw (-2,1) node[anchor=south west] {$+$}
% BATTERY version 1:
%to [battery1] (-3,0) node[anchor=north east] {$-$}
% BATTERY version 2:
-- (-2.45,0.55);
\draw[thick] (-2.85,0.63) -- (-2.05,0.47);
\draw[thick] (-2.75,0.49) -- (-2.35,0.41);
\draw (-2.55,0.45) -- (-3,0) node[anchor=north east] {$-$}
% BATTERY end
to [tline] (2, -1);
% labels
\draw (-3.2,0.8) node {\scalebox{\scl}{$\Delta V$}};
%\draw (3,-0.7) node {$R$}; confusion with Radius, omit!!
% plane front
\draw[densely dotted] (-1.5,-2.75) -- (-1.5,0) -- (1.5,3);
% y and z axis, with part of x also
\draw[darkgray] (0.3,0.3) -- (0.4,0.4);
\draw[darkgray, ->] (0,0) -- (0,1.15) node[above = 5, right = 1] {\scalebox{\scl}{$y$}};
\draw[darkgray, ->] (0,0) -- (1.4,-0.28) node[below = 0.6, right = 1] {\scalebox{\scl}{$z$}};
% current indicators
\draw[->] (-0.4,-0.52) -- (-0.9,-0.42) node[below = 4.3, right = 16] {\scalebox{\scl}{$I$}};
\draw[->] (0.9,0.4) -- (1.4,0.3) node[below = 1.0, right = 1.0] {\scalebox{\scl}{$I$}};
\end{circuitikz}}
\caption{A battery delivering electrical power to a load resistor via two cylindrical wires. The reference Cartesian coordinate system, the current directions and the polarities are shown.\label{fig3d}}
\end{figure}

Consider a battery connected to a load resistor via two infinitely long cylindrical lossless conductors, as shown in Fig.~\ref{fig3d}. The battery has a potential difference~$\Delta V$, and the current in the circuit is~$I$. The two conductors have opposite charge polarities and current directions. The battery is delivering electrical power equal to $I\,\Delta V$, which is absorbed and converted to heat by the resistor. However, how exactly does the electrical power travel from the battery to the resistor in the circuit? The power is shown to be transferred through the electromagnetic field surrounding the conductors~\cite{SEFTON2002,GALILI2005,GRIFFITHS2012,FEYNMAN1964}, as opposed to within them. Yet, the charges in the wires are still required, since they guide the electromagnetic field toward the resistor~\cite{JEFIMENKO1989}.

The following two examples are often used in calculations of the power flowing in the electromagnetic field outside of two current-carrying perfect conductors:

\textit{Example 1}: In the parallel plate DC transmission line~\cite{MORTON1979}, the electric and magnetic fields between the plates are uniform orthogonal fields (when neglecting fringing effects), and the electrical power~$P$ transported by these fields is proved to be equal to $I\,\Delta V$. 

\textit{Example 2}: For the coaxial cable, the same result is also established relatively quickly given its simple symmetry~\cite{GRIFFITHS2012}. % p.368. 
The fields are orthogonal as well, but decrease in $1/r$ away from the center axis, where~$r$ is the distance to the center axis.

In these types of problems, it is possible to analyze the energy as contained in either the fields or the configuration of charges, although in other situations such as general relativity and momentum conservation, the case for considering the fields is much stronger~\cite{GRIFFITHS2012}. % p.97, 361

A pair of parallel wires (or wire pair) is arguably the most ubiquitous 
form of electrical energy transportation. Although a power $I\,\Delta V$ should also be expected to flow in the electromagnetic field of this particular two-conductor geometry, can it be obtained analytically? When the wires are not perfect conductors, how exactly does their resistance affect this power flow? 

Answers to the above questions are presented in this paper, which is organized as follows: 
Sec.~\ref{secAsm} lists the assumptions used in the problem's description, Sec.~\ref{secPrev} presents previous relevant work, Sec.~\ref{secObp} contains a quick review of bipolar cylindrical coordinates (simply called bipolar coordinates hereafter), Secs.~\ref{sec_pec} and~\ref{sec_resist} present the power flow calculations for the perfectly conducting and resistive wire pair, respectively.

%------------------------------------------------------------------------------
\section{Assumptions}
\label{secAsm}
%------------------------------------------------------------------------------
\begin{figure*}
\centerline{\begin{circuitikz}
\def\scl{1.2}
% figure letter
\draw (-3.5,-1.0) node[anchor=north] {(a)};
% axes
\draw[darkgray, ->] (-3.5,0) -- (3.5,0) node[anchor=west] {\scalebox{\scl}{$x$}};
\draw[darkgray, ->] (0,-1.4) -- (0,1.4) node[anchor=south] {\scalebox{\scl}{$y$}};
% coords
\coordinate (AL) at (-1.7,0);
\coordinate (AR) at (1.7,0);
\coordinate (LL) at (-1.85,0);
\coordinate (LR) at (1.85,0);
% big wires
\draw (LL) circle (0.7);
\draw (LR) circle (0.7);
% thin wires
\filldraw[fill=black] (AR) circle (0.06) node[anchor=south west] {\scalebox{\scl}{$\uplambda$}};
\filldraw[fill=black] (AL) circle (0.06) node[anchor=south east] {\scalebox{\scl}{$-\uplambda$}};
% radius R and deltaV/2
\draw (LL) -- +(-135:0.67) node[anchor=north east] {\scalebox{\scl}{$R$}};
\draw (LR) -- +(-45:0.67) node[anchor=north west] {\scalebox{\scl}{$R$}};
\draw (-2.8,0.7) node {\scalebox{\scl}{$-\frac{\Delta V}{2}$}};
\draw (2.7,0.7) node {\scalebox{\scl}{$\frac{\Delta V}{2}$}};
% l markers
\draw[darkgray, <->] (-1.85,-0.9) -- (0,-0.9);
\draw[darkgray, <->] (0,-0.9) -- (1.85,-0.9);
\draw (1,-1.15) node {\scalebox{\scl}{$l$}};
\draw (-1,-1.15) node {\scalebox{\scl}{$l$}};
\draw[darkgray, densely dashed] (-1.85,-0.05) -- (-1.85,-1.1);
\draw[darkgray, densely dashed] (1.85,-0.05) -- (1.85,-1.1);
% a markers
\draw[darkgray, <->] (-1.7,0.9) -- (0,0.9);
\draw[darkgray, <->] (0,0.9) -- (1.7,0.9);
\draw (0.85,0.7) node {\scalebox{\scl}{$a$}};
\draw (-0.85,0.7) node {\scalebox{\scl}{$a$}};
\draw[darkgray, densely dashed] (-1.7,0.1) -- (-1.7,1.1);
\draw[darkgray, densely dashed] (1.7,0.1) -- (1.7,1.1);
% epsilon
\draw (3.3,-1.0) node {\scalebox{\scl}{$\epsilon_0$}};
\end{circuitikz}
\qquad %--------------------------------------------------------------
\begin{circuitikz}
\def\scl{1.2}
% figure letter
\draw (-3.5,-1.0) node[anchor=north] {(b)};
% axes
\draw[darkgray, ->] (-3.5,0) -- (3.5,0) node[anchor=west] {\scalebox{\scl}{$x$}};
\draw[darkgray, ->] (0,-1.4) -- (0,1.4) node[anchor=south] {\scalebox{\scl}{$y$}};
% coords
\coordinate (AL) at (-1.7,0);
\coordinate (AR) at (1.7,0);
\coordinate (LL) at (-1.85,0);
\coordinate (LR) at (1.85,0);
% big wires
\draw (LL) circle (0.7);
\draw (LR) circle (0.7);
% thin wires
\filldraw[fill=white] (AR) circle (0.07) node[anchor=south west] {\scalebox{\scl}{$I$}};
\filldraw[fill=black] (AR) circle (0.02);
\filldraw[fill=white] (AL) circle (0.07) node[anchor=south east] {\scalebox{\scl}{$I$}};
\draw (-1.7424,0.0424) -- (-1.6576,-0.0424);
\draw (-1.7424,-0.0424) -- (-1.6576,0.0424);
% radius R
\draw (LL) -- +(-135:0.67) node[anchor=north east] {\scalebox{\scl}{$R$}};
\draw (LR) -- +(-45:0.67) node[anchor=north west] {\scalebox{\scl}{$R$}};
% l markers
\draw[darkgray, <->] (-1.85,-0.9) -- (0,-0.9);
\draw[darkgray, <->] (0,-0.9) -- (1.85,-0.9);
\draw (1,-1.15) node {\scalebox{\scl}{$l$}};
\draw (-1,-1.15) node {\scalebox{\scl}{$l$}};
\draw[darkgray, densely dashed] (-1.85,-0.05) -- (-1.85,-1.1);
\draw[darkgray, densely dashed] (1.85,-0.05) -- (1.85,-1.1);
% a markers
\draw[darkgray, <->] (-1.7,0.9) -- (0,0.9);
\draw[darkgray, <->] (0,0.9) -- (1.7,0.9);
\draw (0.85,0.7) node {\scalebox{\scl}{$a$}};
\draw (-0.85,0.7) node {\scalebox{\scl}{$a$}};
\draw[darkgray, densely dashed] (-1.7,0.1) -- (-1.7,1.1);
\draw[darkgray, densely dashed] (1.7,0.1) -- (1.7,1.1);
% mu0
\draw (3.3,-1.0) node {\scalebox{\scl}{$\mu_0$}};
\end{circuitikz}
\vspace{-8pt}}
\caption{Cross sections of perfectly conducting thick wires of radii $R$, separated by a distance~$2l$ center-to-center. The polarities (a) and current directions (b) are visible. Also shown are the effective thin wires (at $x=\pm a$) that create equivalent fields to those of the thick wires, outside of the thick wires. All illustrations hereafter use $R=0.7$~cm and $a=1.7$~cm.\label{fig2d}}
\end{figure*}
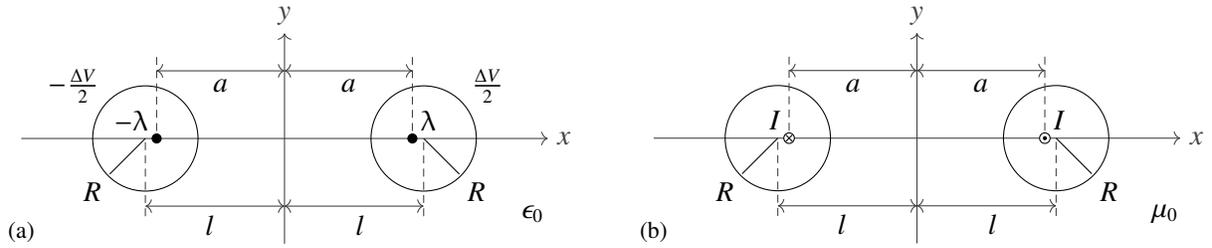

The two wires are linear, parallel, and considered to be infinitely long. The battery and load resistor are sufficiently far away from the region of analysis such that any fringing (or edge) effects are neglected in the~$z$ direction (see Fig.~\ref{fig3d}). 
An infinite vacuum of permittivity~$\epsilon_0$ and permeability~$\mu_0$ surrounds the wires.

When the wires are perfect electrical conductors, they are lossless (i.e., no electrical resistance) and have no initial magnetic fields within them. This implies that (i) inside the wires, the electric field is equal to zero, and the magnetic field remains equal to zero; (ii) just outside of the wires, the electric field is perpendicular to their surfaces and the magnetic field is tangent to their surfaces; and (iii) currents and free charges are located on the surfaces of the wires~\cite{GRIFFITHS2012,PAUL2008,HAUS1989}.
%GRIFFITHS2012: p.296, 297, 346, 425
%PAUL2008: p.18-19, 26, 79, 102, *120*, 139, 182
%HAUS1989: section 8.4, 8.6, 10.0
%more boundary info in Griffiths p.343

When the wires have resistance (i.e., finite conductivity), 
the current density is uniformly distributed in each wire's cross section,
and the potential varies linearly in the~$z$ direction~\cite{GRIFFITHS2012,RUSSELL1968}. Any pinching~\cite{MATZEK1968} or Hall effects are neglected, such that 
the extremely small transverse electric field inside the wires is neglected.
% Griffiths p.299

Resistive wires of infinite length imply a battery of infinite electromotive force (emf). Although the context can be described in this way (see problem IV.2 by Stratton~\cite{STRATTON2007}), it can also be assumed that the wires are instead very long, such that the battery has a finite emf.

%------------------------------------------------------------------------------
\section{Related Work}
\label{secPrev}
%------------------------------------------------------------------------------

The perfectly conducting wire pair has been extensively studied. Using complex analysis, the complex potential of this configuration can be established, whereby the electric scalar potential corresponds to the real part of this complex function~\cite{SCHAUM2009}. The wire pair is also seen as a transmission line, where the electric scalar potential~$V$ and magnetic vector potential~$\vect{A}$ can be determined. These can be used to calculate the capacitance~\cite{POPOVIC1971,HAUS1989} and the inductance of the wire pair~\cite{PAUL2008,HAUS1989}. Multi-conductor cables and asymmetrical conductors have also been explored~\cite{SARMA1969,LEKNER2022}, along with shielded conductors~\cite{NORDGARD1976}, but these will not be considered here. 

The electric equipotentials of a pair of perfectly conducting and oppositely charged wires are circles (or cylinders in three dimensions) but are not centered on the wires' centers. They are instead centered on the locations of two thin wires that create an equivalent field outside of the actual thick wires. By a similar reasoning, the magnetic vector potential of two thin wires carrying opposite currents is shown to be constant over the surfaces of the two thick wires. In both cases, the resulting surface charge and surface current densities can be obtained from the fields.

It can be established that the true centers (at $x=\pm l$) and the effective centers (at $x=\pm a$) of the wires, shown in Fig.~\ref{fig2d}, are related by
\begin{equation}
\label{eqCenters}
a^2=l^2-R^2,
\end{equation}
where~$R$ is the radius of each wire~\cite{PAUL2008,HAUS1989,POPOVIC1971,ENGELEN2013}.

Haus and Melcher~\cite{HAUS1989} showed that the electric scalar and magnetic vector potentials outside and at the surfaces of the perfectly conducting wires are
\begin{equation}
\label{eqV0}
V=\frac{-\uplambda}{2\pi\epsilon_0}\ln\left(\frac{\sqrt{\left(x-a\right)^2+y^2}}{\sqrt{\left(x+a\right)^2+y^2}}\right),% derived from (Haus 4.6.eq18, p.151).
\end{equation}
\begin{equation}
\label{eqvecA0}
\vect{A}=\frac{-\mu_0 I}{2\pi}\ln\left(\frac{\sqrt{\left(x-a\right)^2+y^2}}{\sqrt{\left(x+a\right)^2+y^2}}\right)\hat{\vect{z}},% derived from (Haus 8.6.eq9, p.415)
\end{equation}
where~$\lambda$ is the per unit length charge density of the wires. One wire has a positive charge and the other has a negative charge, as shown in Fig.~\ref{fig2d}.

The most suitable form of the electric potential for use in this paper is actually the one obtained as a solution to problem 3.12 by Griffiths~\cite{GRIFFITHS2012}, or derived from results in Refs.~\onlinecite{ENGELEN2013,HERNANDES2016}:
\begin{equation}
\label{eqV}
V=\frac{\Delta V}{4\arccosh(l/R)}\ln\left(\frac{{\left(x+a\right)^2+y^2}}{{\left(x-a\right)^2+y^2}}\right),
\end{equation}
where the potentials of the wires are~$\pm\Delta V/2$.
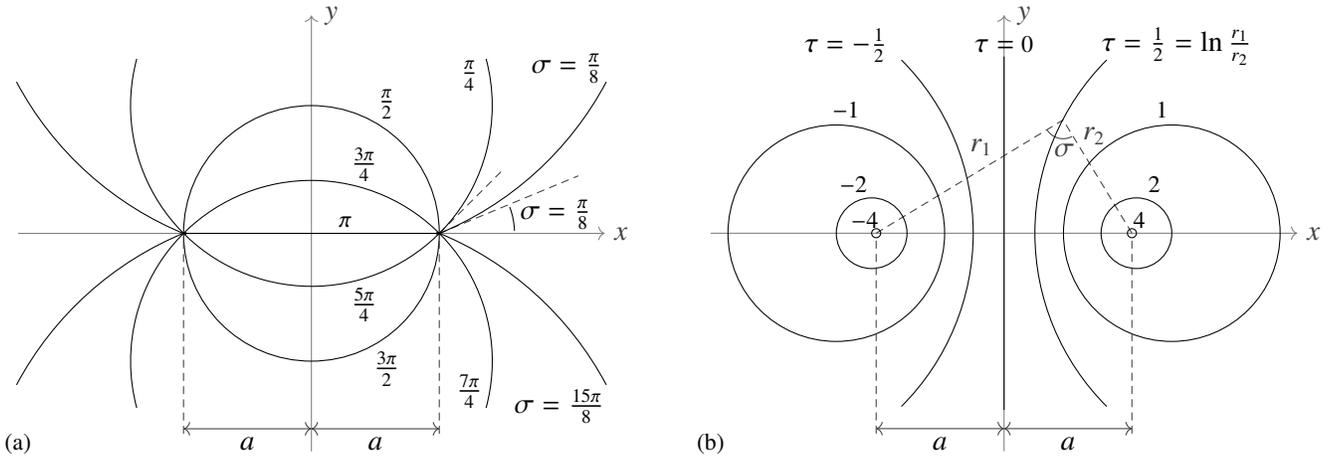
\begin{figure*}% this figure was sized for letterpaper, but if A4 is used, reduce the \qquad in fig3.tex (in between the two circuitikz blocks), and/or use "anchor=north" for the positioning of the "x" axis labels (instead of "anchor=west"). this should be enough to make this figure narrower
\centerline{\begin{circuitikz}
\def\scl{1.2}
% figure letter
\draw (-3.9,-2.8) node {(a)};
% SIGMA
% axes
\draw[gray, ->] (-3.9,0) -- (3.9,0) node[anchor=west,darkgray] {\scalebox{\scl}{$x$}};
\draw[gray, ->] (0,-2.9) -- (0,2.9) node[right = 2,darkgray] {\scalebox{\scl}{$y$}};
% sigma curves
% sigma angle for specific case of pi/8 and pi/4
\draw[darkgray, densely dashed] (1.7,0) -- +(22.5:2);
\draw[darkgray, densely dashed] (1.7,0) -- +(45:1.2);
\centerarc[](1.7,0)(2:20.5:1);
\draw (3.25,0.25) node {\scalebox{\scl}{$\sigma=\frac{\pi}{8}$}};
% sigma = pi/8 and 15pi/8
\centerarc[](0,4.1)(208:248:4.44); \centerarc[](0,4.1)(292:332:4.44);
\centerarc[](0,-4.1)(28:68:4.44); \centerarc[](0,-4.1)(112:152:4.44);
\draw (3.4,2.2) node {\scalebox{\scl}{$\sigma=\frac{\pi}{8}$}};
\draw (3.3,-2.3) node {\scalebox{\scl}{$\sigma=\frac{15\pi}{8}$}};
% sigma = pi/4 and 3pi/4 and 5pi/4 and 7pi/4
\centerarc[](0,1.7)(165:375:2.404);
\centerarc[](0,-1.7)(-15:195:2.404);
\draw (2.1,2.1) node {\scalebox{\scl}{$\frac{\pi}{4}$}};
\draw (0.7,0.92) node {\scalebox{\scl}{$\frac{3\pi}{4}$}};
\draw (0.7,-0.95) node {\scalebox{\scl}{$\frac{5\pi}{4}$}};
\draw (2.1,-2.1) node {\scalebox{\scl}{$\frac{7\pi}{4}$}};
% sigma = pi/2 and 3pi/2
\draw (0,0) circle (1.7);
\draw (1,1.72) node {\scalebox{\scl}{$\frac{\pi}{2}$}};
\draw (1,-1.8) node {\scalebox{\scl}{$\frac{3\pi}{2}$}};
% sigma = pi
\draw (-1.7,0) -- (1.7,0);
\draw (0.45,0.15) node {\scalebox{1}{$\pi$}};
% a markers
\draw[darkgray, <->] (-1.7,-2.6) -- (0,-2.6);
\draw[darkgray, <->] (0,-2.6) -- (1.7,-2.6);
\draw (0.85,-2.8) node {\scalebox{\scl}{$a$}};
\draw (-0.85,-2.8) node {\scalebox{\scl}{$a$}};
\draw[darkgray, densely dashed] (-1.7,-0.1) -- (-1.7,-2.7);
\draw[darkgray, densely dashed] (1.7,-0.1) -- (1.7,-2.7);
\end{circuitikz}
\qquad %--------------------------------------------------------------
\begin{circuitikz}
\def\scl{1.2}
% figure letter
\draw (-3.9,-2.8) node {(b)};
% TAU
% axes
\draw[gray, ->] (-3.9,0) -- (3.9,0) node[anchor=west,darkgray] {\scalebox{\scl}{$x$}};
\draw[gray, ->] (0,-2.9) -- (0,2.9) node[right = 2,darkgray] {\scalebox{\scl}{$y$}};
% tau curves
% tau ln for specific case of 1/2
\coordinate (P) at (0.785,1.507);
\draw[darkgray, densely dashed] (-1.7,0) -- (P);
\draw[darkgray, densely dashed] (1.7,0) -- (P);
\draw[darkgray] (-0.3,1.15) node {\scalebox{\scl}{$r_1$}};
\draw[darkgray] (1.2,1.3) node {\scalebox{\scl}{$r_2$}};
\centerarc[darkgray](P)(212:298:0.25);
\draw[darkgray] (0.8,1.1) node {\scalebox{\scl}{$\sigma$}};
% tau=0 [semithick]
\draw (0.0,-2.35) -- (0.0,2.35);
\draw (0.0,2.51) node {\scalebox{\scl}{$\tau=$}\scalebox{1}{$\;0$}};
% tau=1/2
\draw (0.41,0) arc (0:45:-3.26); \draw (0.41,0) arc (0:-45:-3.26);
\draw (-0.41,0) arc (0:45:3.26); \draw (-0.41,0) arc (0:-45:3.26);
\draw (2.3,2.49) node {\scalebox{\scl}{$\tau=\frac{1}{2}=\ln\frac{r_1}{r_2}$}};
\draw (-2.1,2.5) node {\scalebox{\scl}{$\tau=-\frac{1}{2}$}};
% tau=1
\draw (2.23,0) circle (1.44);
\draw (-2.23,0) circle (1.44);
\draw (2.1,1.65) node {\scalebox{1}{$1$}};
\draw (-2.1,1.65) node {\scalebox{1}{$-1$}};
% tau=2
\draw (1.76,0) circle (0.47);
\draw (-1.76,0) circle (0.47);
\draw (2,0.65) node {\scalebox{1}{$2$}};
\draw (-2,0.65) node {\scalebox{1}{$-2$}};
% tau=4
\draw (1.7,0) circle (0.06);
\draw (-1.7,0) circle (0.06);
\draw (1.8,0.18) node {\scalebox{1}{$4$}};
\draw (-1.85,0.18) node {\scalebox{1}{$-4$}};
% a markers
\draw[darkgray, <->] (-1.7,-2.6) -- (0,-2.6);
\draw[darkgray, <->] (0,-2.6) -- (1.7,-2.6);
\draw (0.85,-2.8) node {\scalebox{\scl}{$a$}};
\draw (-0.85,-2.8) node {\scalebox{\scl}{$a$}};
\draw[darkgray, densely dashed] (-1.7,-0.1) -- (-1.7,-2.7);
\draw[darkgray, densely dashed] (1.7,-0.1) -- (1.7,-2.7);
\end{circuitikz}
\vspace{-4pt}}
\vspace{-2pt}
\caption{Overview of bipolar coordinates. (a) Curves of constant~$\sigma$ are truncated circles that terminate at the focal points. (b) Curves of constant~$\tau$ are non-concentric circles. Any point ($x$, $y$) can be described by its equivalent coordinates ($\sigma$, $\tau$).\label{figBipol}}
\end{figure*} 

Another team has studied the wire pair and reported simulation results for the power transported by opposite direct currents in parallel conductors~\cite{ENGELEN2013}.
The electric potential is calculated based on the Laplace equation along with the boundary conditions of the electric potential on the wires. Although the authors mention treating the case of perfect conductors, the expression used for the magnetic field is actually consistent with that of resistive wires. The analytical expressions of the Poynting vector and the power transported are not given; however, numerical results for fixed wire dimensions are presented. 

In the case of the resistive wire pair, the electric potential and field are calculated by two groups of authors~\cite{HERNANDES2016,ASSIS1999}; however, the magnetic field and Poynting vector are not explored. The Poynting vector has been calculated with the simplifying assumption that $l\gg R$ and in the longitudinal direction only~\cite{MATAR2017}. 

Adapting the electric scalar potential in Eq.~(\ref{eqV}) for resistive wires entails incorporating a linear dependence in~$z$, whereas for the magnetic vector potential in Eq.~(\ref{eqvecA0}), the positions of the thin wires must be the true centers (as opposed to the effective centers). This will be explained in more detail in Sec.~\ref{sec_resist}.

From the potentials, the electric field vector~$\vect{E}$ and the magnetic field vector~$\vect{B}$ in the region surrounding the wires can be calculated using the gradient and curl~\cite{GRIFFITHS2012},
\begin{equation}
\label{eqEphi}
\vect{E}=-\nabla V,% Griffiths p.79
\end{equation}
\begin{equation}
\label{eqBa}
\vect{B}=\vects{\nabla}\times\vect{A}.% Griffiths p.243
\end{equation}
The energy transported by the fields per unit area, per unit time (the power density)  %Griffiths p.358
 is obtained by using the Poynting vector~\cite{GRIFFITHS2012},
\begin{equation}
\label{eqPoy}
\vect{S}=\frac{1}{\mu_0}\left(\vect{E}\times\vect{B}\right).% Griffiths p.358
\end{equation}
The SI units of the Poynting vector~$\vect{S}$ are watts per square meter. When the~$\vect{E}$ and~$\vect{B}$ vectors are in a plane perpendicular to the~$z$ axis, as is the case for perfect conductors, $\vect{S}$ is parallel to the~$z$ axis. For resistive wires, $\vect{S}$ has additional transverse components, which, at the surfaces of the wires, represent power entering the wires from the fields.

The power transported by the electromagnetic field through a given surface can be obtained by integrating the power density over the surface,
\begin{equation}
\label{eqPfromS}
P=\int{\vect{S}\cdot\hat{\vect{n}}\,dA},
\end{equation}
where~$dA$ is an area element (not to be confused with the differential of the magnetic scalar potential), and~$\hat{\vect{n}}$ is a unit vector normal to~$dA$.

%------------------------------------------------------------------------------
\section{Overview of Bipolar Coordinates}
\label{secObp}
%------------------------------------------------------------------------------

Choosing an appropriate coordinate system can ease the solution process when solving electromagnetism problems. 
Calculating the power in the coaxial cable (Example~2, Sec.~\ref{sec_intro}) in Cartesian coordinates leads to intricate integrals, while the calculation in polar coordinates is much less complicated. 
For the wire pair, attempting a solution in Cartesian coordinates leads to integrals with no known primitives. 

Bipolar coordinates~\cite{STRATTON2007,KORN2000} are a type of orthogonal reference system based on two focal points at~$x=\pm a$. %Korn p204 of 1151
An expression in Cartesian coordinates can be transformed into a bipolar one using the following substitutions:
\begin{equation}
\label{eqtx}
x=a\frac{\sinh\tau}{\cosh\tau-\cos\sigma},
\end{equation}
\begin{equation}
\label{eqty}
y=a\frac{\sin\sigma}{\cosh\tau-\cos\sigma}.
\end{equation}
No transformation is needed in $z$. 
The~$\tau$ coordinate ranges from~$-\infty$ at the left focal point to~$+\infty$ at the right focal point. The~$\sigma$ coordinate ranges from~$0$ to~$2\pi$,~\cite{MEHDI2019} although other conventions are possible, providing~$\sigma$ has a range of~$2\pi$. 

As shown in Fig.~\ref{figBipol}, curves of constant~$\sigma$ are non-concentric truncated circles with endpoints located at the two focal points. Curves of constant~$\tau$ are non-concentric circles, with the case $\tau=0$ corresponding to the~$y$ axis. Figure~\ref{figBipol} also illustrates the geometric interpretations of the bipolar variables, namely the two possible (and equivalent) positions of the~$\sigma$ angle, and the definition of~$\tau$ as the natural logarithm of the ratio of the distances to the focal points. 

The scale factors needed to calculate differential operators (such as gradient, divergence, and curl) are
\begin{equation}
\label{eqscst}
h_\sigma = h_\tau = \frac{a}{\cosh\tau-\cos\sigma},
\end{equation}
along with $h_z=1$, when required. The surface element~$dA$ in the $\sigma\tau$-plane becomes
\begin{equation}
\label{eqdAb}
dA = dx\,dy= h_\sigma h_\tau \, d\sigma d\tau,
\end{equation}
and the gradient and curl operators needed in Eqs.~(\ref{eqEphi}) and~(\ref{eqBa}) are calculated using
\begin{equation}
\label{eqgrad}
\nabla F = 
\frac{1}{h_\sigma}\frac{\partial F}{\partial\sigma}\,\hat{\vects{\sigma}}+
\frac{1}{h_\tau}\frac{\partial F}{\partial\tau}\,\hat{\vects{\tau}}+
\frac{1}{h_z}\frac{\partial F}{\partial z}\,\hat{\vect{z}},
\end{equation}
\begin{equation}
\label{eqcurl}
\vects{\nabla}\times\vect{F} = \frac{1}{h_\sigma h_\tau h_z}\begin{vmatrix}
h_\sigma \hat{\vects{\sigma}} & h_\tau \hat{\vects{\tau}} & h_z \hat{\vect{z}}\\ 
\rule{0pt}{13pt}\frac{\partial}{\partial\sigma} & \frac{\partial}{\partial\tau} & \frac{\partial}{\partial z}\\
\rule{0pt}{12pt}h_\sigma F_\sigma & h_\tau F_\tau & h_z F_z
\end{vmatrix},
\end{equation}
where~$F$ is a scalar function of~$\sigma$, $\tau$, and~$z$, and~$\vect{F}$ is a vector function of the same variables, with components~$F_\sigma$, $F_\tau$, and~$F_z$. 
The divergence and Laplacian operators also have formulations in bipolar coordinates, as shown in Refs.~\onlinecite{KORN2000,ELKAMEL2011} and in the online supplementary material~\cite{SUPPMAT}, but they will not be used here.

The cross product and dot product are identical to those used in Cartesian coordinates, and are given by
\begin{equation}
\label{eqcross}
\vect{F}\times\vect{G} = \begin{vmatrix}
\hat{\vects{\sigma}} & \hat{\vects{\tau}} & \hat{\vect{z}}\\ 
F_\sigma & F_\tau & F_z\\
G_\sigma & G_\tau & G_z
\end{vmatrix},
\end{equation}
\begin{equation}
\label{eqdot}
\vect{F}\cdot\vect{G} =  
F_\sigma G_\sigma + F_\tau G_\tau + F_z G_z.
\end{equation}

An extensive list of coordinate systems along with their scale factors and differential operators is found in the work of Elkamel \textit{et al}. on partial differential equations~\cite{ELKAMEL2011}.

%------------------------------------------------------------------------------
\section{Perfectly Conducting Wires}
\label{sec_pec}
%------------------------------------------------------------------------------

The objective in this section is to demonstrate that the surface integral of the Poynting vector over a transverse plane outside of the perfectly conducting wires is equal to $I\,\Delta V$.
The~$\tau$ values of the wires must first be established. The surface of the right wire is located at $\tau=\tau_w$, while the left wire is located at $\tau=-\tau_w$. An expression for the coordinate~$\tau_w$ has been previously calculated~\cite{ENGELEN2013}, 
\begin{equation}
\label{eqtw}
\tau_w=\arccosh(l/R),
\end{equation}
but it can be obtained by setting $V=\Delta V/2$ and $\tau=\tau_w$ in Eq.~(\ref{eqVb}) and solving for~$\tau_w$.

\subsection{Potentials and Fields}
\label{ssecPF}
%------------------------------------------------------------------------------

The potentials are much simpler in bipolar coordinates. Applying the substitutions in Eqs.~(\ref{eqtx}) and~(\ref{eqty}) to Eqs.~(\ref{eqvecA0}) and~(\ref{eqV}) yields
\begin{equation}
\label{eqVb}
V=\frac{\Delta V}{2\arccosh(l/R)}\tau,
\end{equation}
\begin{equation}
\label{eqvecAb}
\vect{A}=\frac{\mu_0 I}{2\pi}\tau\,\hat{\vect{z}}.
\end{equation}
The fact that both expressions are functions only of~$\tau$ shows that curves of constant~$\tau$ correspond to curves of constant electric and magnetic potentials. 

The gradient and curl operators in Eqs.~(\ref{eqgrad}) and~(\ref{eqcurl}) can be simplified given that $\partial V/\partial\sigma$, $\partial V/\partial z$, $A_\sigma$, $A_\tau$, $\partial A_z/\partial\sigma$, and $\partial A_z/\partial z$ are all equal to zero, and~$h_z=1$
\begin{equation}
\label{eqgrad2}
\vect{E}=-\nabla V = -\frac{1}{h_\tau}\frac{\partial V}{\partial\tau}\,\hat{\vects{\tau}},
\end{equation}
\begin{equation}
\label{eqcurl2}
\vect{B}=\vects{\nabla}\times\vect{A} = \frac{1}{h_\tau}\frac{\partial A_z}{\partial\tau}\,\hat{\vects{\sigma}}.
\end{equation}

The electric and magnetic fields can be obtained by performing the above-mentioned calculations using the potentials in Eqs.~(\ref{eqVb}) and~(\ref{eqvecAb}), resulting in
\begin{equation}
\label{eqEb}
\vect{E}=-\frac{(\cosh\tau-\cos\sigma)\Delta V}{2a\arccosh(l/R)}\,\hat{\vects{\tau}},
\end{equation}
\begin{equation}
\label{eqBb}
\vect{B}=\frac{(\cosh\tau-\cos\sigma)\mu_0 I}{2\pi a}\,\hat{\vects{\sigma}}.
\end{equation}
\begin{figure}% only located here in the tex for optimal placement in the pdf
\centerline{\begin{circuitikz}
\def\scl{1.2}
% coords
\coordinate (LL) at (-1.85,0);
\coordinate (LR) at (1.85,0);
% big wires
\draw[fill=lightgray] (LL) circle (0.7);
\draw[fill=lightgray] (LR) circle (0.7);
% dA
\coordinate (DABOT) at (-0.785,-1.507);
\coordinate (DALEFT) at (-0.907,-1.438);
\coordinate (DATOP) at (-0.836,-1.319);
\coordinate (DARIGHT) at (-0.722,-1.378);
\filldraw[black] (DABOT) -- (DALEFT) -- (DATOP) -- (DARIGHT) -- cycle;
\centerarc[gray](-3.187,0)(315:342:2.695);% tau=-0.595
\centerarc[dashed,gray](0,0.1705)(220:260:1.7085);% sigma=3Pi/2-0.1
\draw (-0.6,-1.9) node {\scalebox{\scl}{$dA$}};
% SIGMA
% sigma = pi/8 and 15pi/8
\centerarc[dashed,gray](0,4.1)(205:248:4.44); \centerarc[dashed,gray](0,4.1)(292:335:4.44);
\centerarc[dashed,gray](0,-4.1)(25:68:4.44); \centerarc[dashed,gray](0,-4.1)(112:155:4.44);
% sigma = pi/4 and 3pi/4 and 5pi/4 and 7pi/4
\centerarc[dashed,gray](0,1.7)(165:375:2.404);
\centerarc[dashed,gray](0,-1.7)(-15:195:2.404);
% sigma = pi/2 and 3pi/2
\centerarc[dashed](0,0)(0:180:1.7);
\centerarc[dashed,gray](0,0)(180:360:1.7);
\draw (-0.03,1.4) node {\scalebox{\scl}{$\sigma=\frac{\pi}{2}$}};
% sigma = pi
\draw [dashed,gray] (-1.7,0) -- (1.7,0);
% TAU
% tau=0
\draw [gray] (0,-2.35) -- (0,2.35);
% tau=1/2
\draw (0.41,0) arc (0:45:-3.26); \draw (0.41,0) arc (0:-45:-3.26);
\draw [gray] (-0.41,0) arc (0:45:3.26); \draw [gray] (-0.41,0) arc (0:-45:3.26);
\draw (1.6,2.45) node {\scalebox{\scl}{$\tau=\frac{1}{2}$}};
% tau=1
\draw [gray] (2.23,0) circle (1.44);
\draw [gray] (-2.23,0) circle (1.44);
\draw (3.95,0.8) node {\scalebox{\scl}{$\tau\!=\!1$}};
% tau=2
\draw [gray] (1.76,0) circle (0.47);
\draw [gray] (-1.76,0) circle (0.47);
% tau=4
\draw [gray] (1.7,0) circle (0.06);
\draw [gray] (-1.7,0) circle (0.06);
\coordinate (P) at (0.785,1.507);
\draw (P) node[fill,circle,inner sep=0pt,minimum size=3pt] {};
% unit vectors
\draw[-latex,thick] (P) -- +(-27.5:0.5); \draw (1.3,1.5) node {\scalebox{\scl}{$\hat{\vects{\tau}}$}};
\draw[-latex,thick] (P) -- +(-117.5:0.5); \draw (0.8,1.05) node {\scalebox{\scl}{$\hat{\vects{\sigma}}$}};
% perp
\draw[] (P) -- +(-27.5:0.14) -- +(-72.5:0.2) -- +(-117.5:0.14);

% E and B vectors
\draw[-latex,thick] (P) -- +(152.5:0.9); 
\draw (0.19,2.12) node {\scalebox{\scl}{$\vect{E}$}};
\draw[-latex,thick] (P) -- +(-117.5:0.9); 
\draw (0.17,0.88) node {\scalebox{\scl}{$\vect{B}$}};

% tw
\draw (2.93,0.15) node {\scalebox{\scl}{$\tau_w{\scriptstyle\approx}$}};
\draw (2.93,-0.15) node {\scalebox{1}{$1.62$}};
\draw (-2.95,0) node {\scalebox{\scl}{$-\tau_w$}};
\end{circuitikz}
\vspace{-4pt}}
\caption{Electric and magnetic fields at the point given by the coordinates $(\sigma=\frac{\pi}{2}, \tau=\frac{1}{2})$. Also visible are the~$\tau$ values of the wires ($\pm\tau_w$) and the area element~$dA$ used in the power integral. All unlabeled curves are consistent with those shown in Figure~\ref{figBipol}.\label{figBipol2}}
\end{figure}
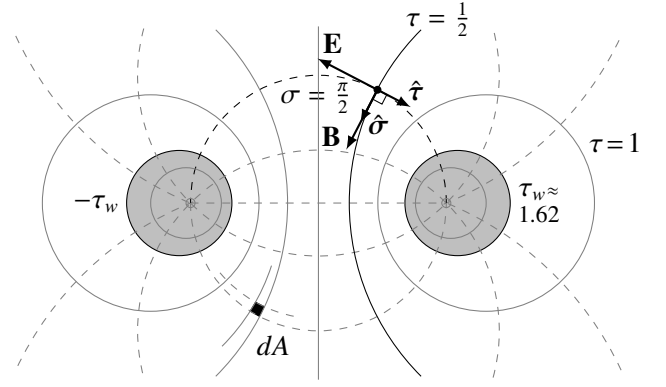

The electric field is described by a single negative component in~$\hat{\vects{\tau}}$ since the electric field is normal to the equipotentials (curves of constant~$\tau$) and is directed from the right wire toward the left wire, which is opposite to the direction of increasing~$\tau$. 
The magnetic field is described by a single positive component in~$\hat{\vects{\sigma}}$. 
The~$\vect{E}$ and~$\vect{B}$ vectors are, therefore, always orthogonal, and are perpendicular and tangent to the surfaces of the wires, respectively, which is consistent with the boundary conditions of perfect conductors presented in Sec.~\ref{secAsm}.
\begin{figure*}
\centerline{\includegraphics[width=0.9\linewidth]{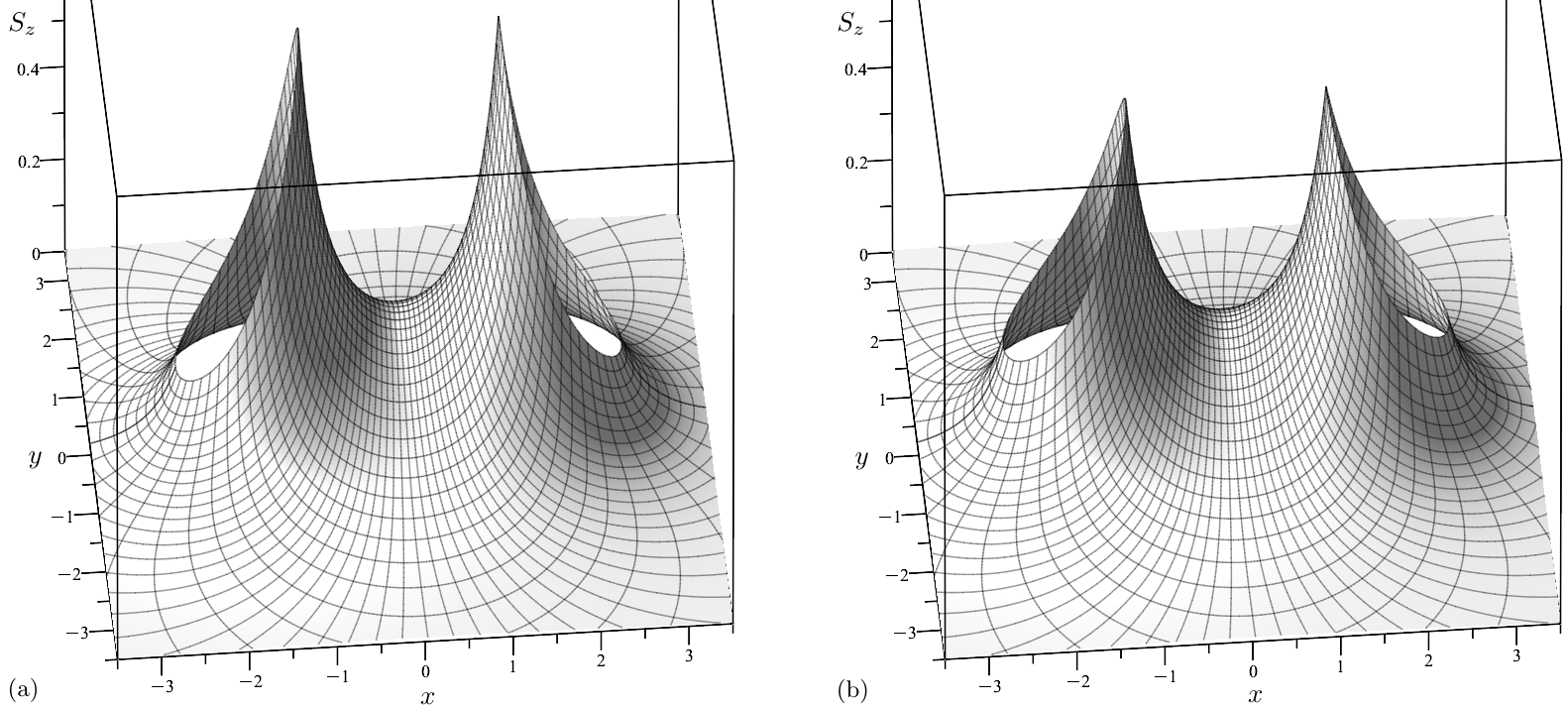}}
\vspace{-4pt}
\caption{The $z$ component of the Poynting vector in the space surrounding the wires, represented by the two holes, for (a) perfectly conducting wires, and (b) resistive wires at $z=0$. In both cases, the maximum power density is located at the inner portions of the wires' cross sections, and power flows towards the resistor around both wires even though they carry oppositely directed currents. The illustration parameters are: $\Delta V=4$~V and $I=1$~A, with $x$ and $y$ in~cm, and $S_z$ in $\mathrm{W/cm}^2$.
\label{figPoy3D}}
\end{figure*}

The apparent simplicity in the~$\vect{E}$ and~$\vect{B}$ vectors is attributable to the choice of an optimal coordinate system: comparing the $\ln()$ appearing in Fig.~\ref{figBipol}(b) with those in Eqs.~(\ref{eqV0}) and (\ref{eqvecA0}) reveals why. Figure~\ref{figBipol2} shows the vector fields at an arbitrary point in the region surrounding the wires. 
The~$\tau$ values of the wires ($\pm\tau_w$) are also visible.

\subsection{The Electromagnetic Power}
%------------------------------------------------------------------------------

With the fields in Eqs.~(\ref{eqEb}) and~(\ref{eqBb}), the Poynting vector becomes
\begin{equation}
\vect{S}=
\frac{1}{\mu_0}\left(\vect{E}\times\vect{B}\right)=\frac{-E_\tau B_\sigma}{\mu_0}\,\hat{\vect{z}},
\end{equation}
which is equal to
\begin{equation}
\label{eqPoy2b}
\vect{S}=\frac{(\cosh\tau-\cos\sigma)^2I\,\Delta V}{4\pi a^2\arccosh(l/R)}\,\hat{\vect{z}}.
\end{equation}
The Poynting vector has a single~$z$ component, called~$S_z$. Since $S_z>0$, the power flows toward the resistor everywhere outside of the wires. The graph of $S_z$ is shown in Fig.~\ref{figPoy3D}(a) as a function of~$x$ and~$y$. 

The longitudinal power transported by the electromagnetic field can be obtained by integrating~$S_z$ over a plane of constant~$z$ in the region outside of the wires (i.e., from their surfaces to infinity). If~$dA$ is an area element in this plane (shown in Fig.~\ref{figBipol2}), the power integral reduces to
\begin{equation}
\label{eqPzfromSz}
P_z=\int{\vect{S}\cdot\hat{\vect{z}}\,dA}=\int{S_z\,dA}.
\end{equation}
Using the surface element in Eq.~(\ref{eqdAb}), the power calculation becomes 
\begin{equation}
\label{eqInt}
P_z=\int_{-\tau_w}^{\tau_w}\int_{0}^{2\pi}
S_z\, h_\sigma h_\tau \, d\sigma d\tau.
\end{equation}
Inserting the~$z$ component of the Poynting vector~$\vect{S}$ from Eq.~(\ref{eqPoy2b}) and the scale factors from Eq.~(\ref{eqscst}), without simplifying the resulting expression, the power~$P_z$ is obtained by the calculation of
\begin{equation}
\label{eqInt2b}
\int\limits_{-\tau_w}^{\tau_w}
\int\limits_{0}^{2\pi}
\frac{(\cosh\tau-\cos\sigma)^2 I\,\Delta V a^2}{4\pi a^2\arccosh(l/R)(\cosh\tau-\cos\sigma)^2}
\, d\sigma d\tau.
\end{equation}
This is a double definite integral of a constant since the expressions in~$\tau$ and~$\sigma$ cancel out. When simplified, and when using the~$\tau_w$ value from Eq.~(\ref{eqtw}) in the bounds, the double integral is
\begin{equation}
\label{eqInt2}
P_z=\frac{I\,\Delta V}{4\pi\arccosh(l/R)}\int\limits_{-\arccosh(l/R)}^{\arccosh(l/R)}
\int\limits_{0}^{2\pi}
\, d\sigma d\tau.
\end{equation}
When evaluated, the power transported by the electromagnetic field surrounding the wires is found to be
\begin{equation}
\label{eqP}
P_z=I\,\Delta V,
\end{equation}
as expected.

%------------------------------------------------------------------------------
\subsection{Surface Current and Surface Charge Densities}
\label{ssecDens}
%------------------------------------------------------------------------------

The currents and free charges located on the surfaces of the wires are not uniformly distributed around the circular cross sections of the wires. This is illustrated
on p.~125 of Ref.~\onlinecite{PAUL2008} and p.~344 of Ref.~\onlinecite{HAUS1989}, where these densities are shown to be stronger on the inner portions of the wires' cross sections, although their analytic forms are not calculated.

The fields at the surfaces of the conductors are related to the current and free charge densities according to%~\cite{HAUS1989}
\begin{equation}
\label{eqFiolE}
\hat{\vect{n}}\cdot\vect{E}_w=\frac{\sigc}{\epsilon_0},
\end{equation}
\begin{equation}
\label{eqFiolB}
\hat{\vect{n}}\times\vect{B}_w=\mu_0\vect{K},
\end{equation}
where~$\sigc$ is used in place of the usual surface charge density~$\sigma$ to avoid any confusion with the bipolar coordinate~$\sigma$. The surface current density vector is denoted as~$\vect{K}$, and $\hat{\vect{n}}$ is a unit vector perpendicular to the surface of the wires pointing outward.
For the right wire, the vector~$\hat{\vect{n}}$ is directly related to the unit vector~$\hat{\vects{\tau}}$ by
\begin{equation}
\label{eqnt}
\hat{\vect{n}} = -\hat{\vects{\tau}}.
\end{equation}

The electric and magnetic fields at the surface of the right wire are calculated by evaluating Eqs.~(\ref{eqEb}) and~(\ref{eqBb}) at the~$\tau_w$ position from Eq.~(\ref{eqtw}), resulting in
\begin{equation}
\label{eqEbB}
\vect{E}_w=-\frac{\left(l/R - \cos\sigma\right)\Delta V}{2a\arccosh(l/R)}\,\hat{\vects{\tau}},
\end{equation}
\begin{equation}
\label{eqBbB}
\vect{B}_w=\frac{\left(l/R - \cos\sigma\right)\mu_0 I}{2\pi a}\,\hat{\vects{\sigma}}.
\end{equation}

From Eqs.~(\ref{eqFiolE})--(\ref{eqBbB}), the surface charge and surface current densities of the right wire are found to be
\begin{equation}
\label{eqSig}
\sigc=\frac{\left(l/R - \cos\sigma\right)\epsilon_0 \Delta V}{2\arccosh(l/R)\sqrt{l^2-R^2}},
\end{equation}
\begin{equation}
\label{eqK}
\vect{K}=\frac{\left(l/R - \cos\sigma\right)I}{2\pi\sqrt{l^2-R^2}}\,\hat{\vect{z}},
\end{equation}
where Eq.~(\ref{eqCenters}) has been used to express these as functions of~$l$ instead of~$a$.

It can be shown that the following equality allows the densities to be expressed using a polar angle~$\theta$:
\begin{equation}
\label{eqsigtheta}
\cos\sigma = \frac{R+l\cos\theta}{l + R\cos\theta},
\end{equation}
where $\theta=\pi$ is the leftmost point of the right wire's cross section, and~$\theta=0$ (or~$2\pi$) is the rightmost point of the right wire's cross section. 
The surface charge and surface current densities on the right wire then become
\begin{equation}
\label{eqSig2}
\sigc=\frac{\epsilon_0\Delta V\sqrt{l^2-R^2}}{2R\arccosh(l/R)\,(l+R\cos\theta)},
\end{equation}
\begin{equation}
\label{eqK2}
\vect{K}=\frac{I\sqrt{l^2-R^2}}{2\pi R\,(l+R\cos\theta)}\,\hat{\vect{z}}.
\end{equation}

These densities are strictly positive (since~$l>R$) and are consistent with the polarity and current direction of the right wire shown in Fig.~\ref{fig2d}. The densities have maximum values at~$\theta=\pi$ and minimum values at~$\theta=0$ (or~$2\pi$). The densities of the left wire are $-\sigc$ and $-\vect{K}$.

Making use of trigonometric identities, the charge density in Eq.~(\ref{eqSig2}) can be shown to be equivalent to the one reported in Ref.~\onlinecite{ENGELEN2013}.

%------------------------------------------------------------------------------
\section{Resistive Wires}
\label{sec_resist}
%------------------------------------------------------------------------------

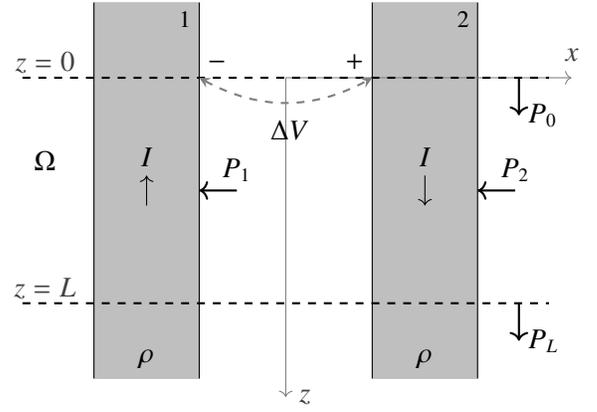
\begin{figure}
\centerline{\begin{circuitikz}
\def\scl{1.2}
% wires
\filldraw [fill=lightgray, draw=lightgray] (-2.55,-1) rectangle (-1.15,4) node[anchor=north east] {$1$};
\draw[black] (-2.55,4) -- (-2.55,-1);
\draw[black] (-1.15,4) -- (-1.15,-1);
\filldraw [fill=lightgray, draw=lightgray] (1.15,-1) rectangle (2.55,4) node[anchor=north east] {$2$};
\draw[black] (2.55,4) -- (2.55,-1);
\draw[black] (1.15,4) -- (1.15,-1);
% axes
\draw[gray, ->] (0,3) -- (0,-1.25) node[right = 2,darkgray] {\scalebox{\scl}{$z$}};
\draw[gray, ->] (0,3) -- (3.8,3) node[above = 3,darkgray] {\scalebox{\scl}{$x$}};
%z planes
\draw[dashed,thick] (-3.5,0) -- (3.5,0);
\draw[dashed,thick] (-3.5,3) -- (3.5,3);
% currents
\draw[->] (-1.85,1.3) -- (-1.85,1.7) node[above = 0.2] {\scalebox{\scl}{$I$}};
\draw[<-] (1.85,1.3) -- (1.85,1.7) node[above = 0.2] {\scalebox{\scl}{$I$}};
% L
\draw[darkgray] (-3.2,0.2) node {\scalebox{\scl}{$z=L$}};
\draw[darkgray] (-3.2,3.2) node {\scalebox{\scl}{$z=0$}};
\draw[] (-3.2,1.9) node {\scalebox{\scl}{$\Omega$}};
% power
\draw[->,thick] (3.1,3) -- (3.1,2.5) node[anchor=west] {\scalebox{\scl}{$P_0$}};
\draw[->,thick] (3.1,0) -- (3.1,-0.5) node[anchor=west] {\scalebox{\scl}{$P_L$}};
\draw[<-,thick] (-1.15,1.5) -- (-0.65,1.5) node[anchor=south] {\scalebox{\scl}{$P_1$}};
\draw[<-,thick] (2.55,1.5) -- (3.05,1.5) node[anchor=south] {\scalebox{\scl}{$P_2$}};
% delta V0
\coordinate (CTL) at (-1.15,3);
\coordinate (CTR) at (1.15,3);
\draw [stealth-stealth,thick,dashed,gray] (CTL.north) to [out=-30,in=210] (CTR.north);
\draw[] (0.05,2.3) node {\scalebox{\scl}{$\Delta V$}};
\draw[] (CTL) node[anchor=south west] {\scalebox{\scl}{$-$}};
\draw[] (CTR) node[anchor=south east] {\scalebox{\scl}{$+$}};
% delta VL
\coordinate (CBL) at (-1.15,0);
\coordinate (CBR) at (1.15,0);
% Rho
\draw[] (-1.85,-0.5) node[anchor=north] {\scalebox{\scl}{$\rho$}};
\draw[] (1.85,-0.5) node[anchor=north] {\scalebox{\scl}{$\rho$}};
\end{circuitikz}}
\vspace{-4pt}
\caption{Two long resistive wires and the region of interest~$\Omega$, where the conservation of energy will be verified with the equation $P_0 = P_L+P_1+P_2$.\label{figResistance}}
\end{figure}

The analysis of the resistive wires is performed in a region~$\Omega$ that extends outward to infinity in the radial direction but is bounded by the two planes at~$z=0$ and~$z=L$ and by the surfaces of the wires, as shown in Fig.~\ref{figResistance}. The objective is to show that the power entering the region at the~$z=0$ plane is equal to the sum of the power leaving the region at~$z=L$ and the power entering both wires. 
It is also an implicit goal to confirm that the power entering the region is still given by~$P=I\,\Delta V$, where~$\Delta V$ is now defined as the potential difference of the wires at $z=0$. 

\subsection{Potentials and Fields}
\label{ssecPFr}
%------------------------------------------------------------------------------

The electric potential of the resistive wires is a linear function of $z$, and the current density~$J$ is uniform inside the wires~\cite{GRIFFITHS2012,RUSSELL1968,CHABAY2020}. The potential difference between the wires can be adjusted with the help of Ohm's law,
\begin{equation}
\label{eqdVz}
\Delta V(z) = \Delta V - 2E_\ell z = \Delta V - 2\rho J z,
\end{equation}
where the longitudinal electric field inside the wires is~$E_\ell$, and the resistivity of the wires is~$\rho$. Boundary conditions~\cite{GRIFFITHS2012} require that this field is also equal to the longitudinal field just outside of the wires. 
% Griffiths p.343-344

The electric potential~$V$ outside of the resistive wires is similar to the one used for perfect conductors in Eq.~(\ref{eqV}), with the exception that the potential difference $\Delta V$ is now $\Delta V(z)$ from Eq.~(\ref{eqdVz}), such that~$V$ becomes
\begin{equation}
\label{eqVr}
V=\frac{\Delta V-2\rho J z}{4\arccosh(l/R)}\ln\left(\frac{{\left(x+a\right)^2+y^2}}{{\left(x-a\right)^2+y^2}}\right).
\end{equation}

As for the magnetic vector potential $\vect{A}$ outside of the resistive wires, the positions of the thin wires in Eq.~(\ref{eqvecA0}) are now located at the true centers $x=\pm l$ instead of the effective centers $x=\pm a$, resulting in
\begin{equation}
\label{eqvecAr0}
\vect{A}=\frac{-\mu_0 I}{2\pi}\ln\left(\frac{\sqrt{\left(x-l\right)^2+y^2}}{\sqrt{\left(x+l\right)^2+y^2}}\right)\hat{\vect{z}}.
\end{equation}
This can be confirmed by using Amp\`ere's law and superposition, and showing that the~$\vect{B}$ field obtained is identical to the curl of~$\vect{A}$ above. As before, $l$ is related to~$a$ and~$R$ using Eq.~(\ref{eqCenters}).

An equivalent form of Eq.~(\ref{eqvecAr0}) can be shown to be
\begin{equation}
\label{eqvecAr}
\vect{A}=\frac{\mu_0 I}{2\pi}\arctanh\!\left(\frac{2lx}{x^2+y^2+l^2}\right)\hat{\vect{z}}.
\end{equation}

By applying the bipolar substitutions for~$x$ and~$y$ outlined in Eqs.~(\ref{eqtx}) and (\ref{eqty}), the electric scalar potential and magnetic vector potential in Eqs.~(\ref{eqVr}) and~(\ref{eqvecAr}) become
\begin{equation}
\label{eqVbr}
V=\frac{(\Delta V-2\rho J z)\tau}{2\arccosh(l/R)},
\end{equation}
\begin{equation}
\label{eqvecAbr}
\vect{A}=\frac{\mu_0 I}{2\pi}\arctanh\!\left(\frac{2k\sinh\tau}{(k^2+1)\cosh\tau+(k^2-1)\cos\sigma}\right)\hat{\vect{z}},
\end{equation}
where the change of variables
\begin{equation}
\label{eqSubsk}
k=\frac{a}{l}
\end{equation}
is used in $\vect{A}$ to simplify further calculations.

The gradient and curl operators in Eqs.~(\ref{eqgrad}) and~(\ref{eqcurl}) can be simplified
given that $\partial V/\partial\sigma$, $\partial A_z/\partial z$, $A_\sigma$, and $A_\tau$ are all equal to zero, and $h_z=1$
\begin{equation}
\label{eqgrad2r}
\vect{E}=-\nabla V = -\frac{1}{h_\tau}\frac{\partial V}{\partial\tau}\hat{\vects{\tau}} -\frac{\partial V}{\partial z}\hat{\vect{z}},
\end{equation}
\begin{equation}
\label{eqcurl2r}
\vect{B}=\vects{\nabla}\times\vect{A} = \frac{1}{h_\tau}\frac{\partial A_z}{\partial\tau}\hat{\vects{\sigma}} -\frac{1}{h_\sigma}\frac{\partial A_z}{\partial\sigma}\hat{\vects{\tau}}.
\end{equation}

Performing the above-mentioned calculations using the potentials in Eqs.~(\ref{eqVbr}) and~(\ref{eqvecAbr}) leads to
\begin{equation}
\label{eqEbr}
\vect{E}=-\frac{\Delta V-2\rho J z}{2h_\tau\arccosh(l/R)}\,\hat{\vects{\tau}} + \frac{\rho J\tau}{\arccosh(l/R)}\,\hat{\vect{z}},
\end{equation}
\begin{equation}
\label{eqBbr}
\vect{B}=\frac{\mu_0 I}{2\pi h_\tau}\beta\,\hat{\vects{\sigma}} -
         \frac{\mu_0 I}{\pi h_\sigma}\gamma\,\hat{\vects{\tau}},
\end{equation}
with
\begin{equation}
\label{eqBeta}
\beta=\frac{2\left(k^{2}+1+\left(k^{2}-1\right) \cosh\tau \cos\sigma\right)k}
{\left((k^{2}+1) \cosh\tau+(k^{2}-1)\cos\sigma \right)^{2}-4 k^{2} \sinh^2\tau},
\end{equation}
and
\begin{equation}
\label{eqGamma}
\gamma=\frac{\left(k^{2}-1\right) k\sinh\tau \sin\sigma}
{\left((k^{2}+1) \cosh\tau+(k^{2}-1)\cos\sigma \right)^{2}-4 k^{2} \sinh^{2}\tau},
\end{equation}
and the scale factors as defined in Eq.~(\ref{eqscst}). 

When compared to the fields of the perfect conductors in the previous section, the electric field now has an additional~$z$ component, which is positive around the right wire and negative around the left wire, and the magnetic field has an additional $\tau$ component and is no longer tangential at the wires' surfaces.

\subsection{The Poynting Vector}
\label{ssecPoyr}
%------------------------------------------------------------------------------

With the fields in Eqs.~(\ref{eqEbr}) and~(\ref{eqBbr}), the Poynting vector in the space surrounding the resistive wires is
\begin{equation}
\label{eqPoy2r}
\vect{S}=
\frac{1}{\mu_0}\left(\vect{E}\times\vect{B}\right)=
-\frac{E_z B_\tau}{\mu_0}\hat{\vects{\sigma}}
+\frac{E_z B_\sigma}{\mu_0}\hat{\vects{\tau}}
-\frac{E_\tau B_\sigma}{\mu_0}\hat{\vect{z}},
\end{equation}
which is equal to
\begin{equation}
\label{eqPoy2br}
\vect{S} = \frac{I}{\pi\arccosh(l/R)}\bigg(\frac{\rho J \tau}{h_\sigma}\,\gamma\,\hat{\vects{\sigma}}\,+\frac{\rho J \tau}{2 h_\tau}\,\beta\,\hat{\vects{\tau}}\,+
           \frac{\Delta V-2\rho J z}{4 (h_\tau)^2}\,\beta\,\hat{\vect{z}}\bigg).
\end{equation}

The $z$ component of the Poynting vector is shown in Fig.~\ref{figPoy3D}(b), in the plane $z=0$. The electrical power is slightly more evenly distributed around the wires' cross sections, when compared to the perfect conductors. The electric field is purely longitudinal inside the resistive wires; thus, no power flows in the~$z$ direction within them.

\begin{figure}
\centerline{\includegraphics[width=3.35in]{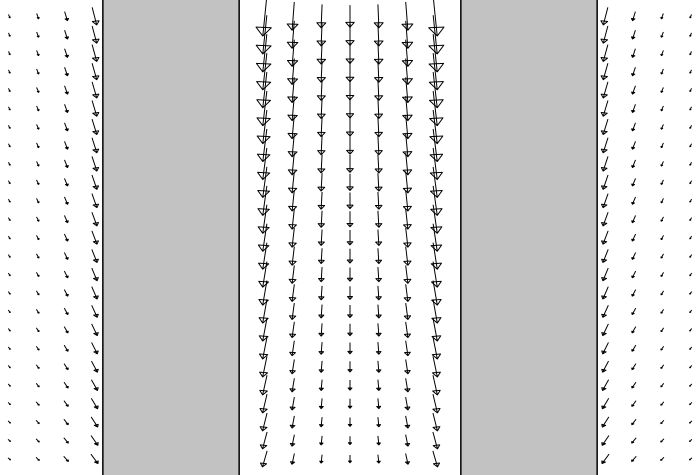}}
\caption{Poynting vector field outside of the resistive wires in the plane~$y=0$, similar to the view in Fig.~\ref{figResistance}. The direction of increasing~$z$ is downwards (towards the load resistor).\label{figSField}}
\end{figure}

Figure~\ref{figSField} shows the Poynting vector field in the plane $y=0$. 
The direction of increasing~$z$ is downward, toward the resistor.
The longitudinal power decreases with increasing~$z$ since a portion of the power is entering the wires. The transverse (horizontal) component of the Poynting vector is constant in~$z$, but is greater near the interior portions of the wires at $x=\pm(l-R)$, when compared to the outer portions at $x=\pm(l+R)$.

\begin{figure}[b]
\centerline{\includegraphics[width=3.35in]{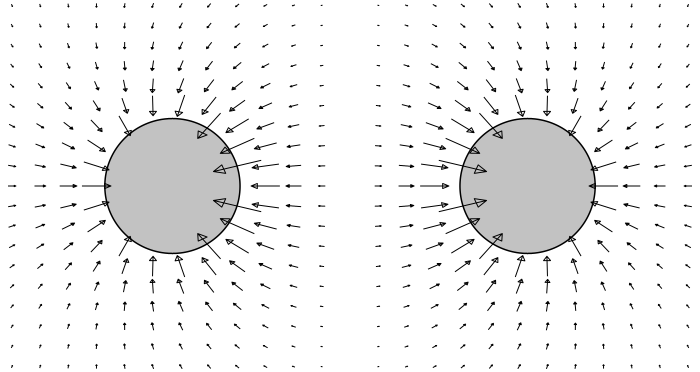}}
\caption{Poynting vector field outside of the resistive wires in the plane~$z=0$ (using a different scale than in Figure~\ref{figSField}).\label{figSField2}}
\end{figure}

Figure~\ref{figSField2} shows the Poynting vector field in the plane $z=0$, where it can also be observed that power enters the wires at the surfaces, but more so in their interior portions.

\subsection{The Longitudinal Power}
\label{ssecPz}
%------------------------------------------------------------------------------

The power $P_z$ flowing parallel to the resistive wires, in the space surrounding the wires, can be calculated in a similar manner as in Eq.~(\ref{eqInt}), namely,
\begin{equation}
\label{eqIntr}
P_z=\int{S_z\,dA}=\int_{-\tau_w}^{\tau_w}\int_{0}^{2\pi}
S_z\, h_\sigma h_\tau \, d\sigma d\tau.
\end{equation}
Since~$h_\sigma=h_\tau$, both scale factors cancel out with the ones in~$S_z$. 
The inner integral reduces to
\begin{equation}
\label{eqIntzsr}
\frac{(\Delta V-2\rho J z)I}{4\pi\arccosh(l/R)}\left(\int_{0}^{2\pi}\beta\, d\sigma\right),
\end{equation}
after a constant factor is moved out of the integral. The integrand is a function of~$\sigma$ and~$\tau$. The integral in parentheses above requires particular attention because of branching, and is solved in the appendix, where it is shown to be equal to~$2\pi$. The inner integral evaluates to 
\begin{equation}
\label{eqIntzs2r}
\frac{(\Delta V-2\rho J z)I}{2\arccosh(l/R)}.
\end{equation}
The outer integral is now evaluated,
\begin{equation}
\label{eqIntzrr}
P_z=\int\limits_{-\arccosh(l/R)}^{\arccosh(l/R)}\frac{(\Delta V-2\rho J z)I}{2\arccosh(l/R)}\, d\tau
= (\Delta V-2\rho J z)I.
\end{equation}
Using the usual definition of the current density~$J$, the power flowing through a transverse plane at position~$z$, outside of the wires, is therefore
\begin{equation}
\label{eqPzr}
P_z=I\,\Delta V-2\frac{\rho z}{\pi R^2}I^2.
\end{equation}
The radius~$R$ of the wires is not to be confused with the resistance of each wire over the distance~$z$, which is actually the fraction appearing in Eq.~(\ref{eqPzr}).

It is therefore shown that the longitudinal power entering and leaving the region of interest~$\Omega$ are, respectively,
\begin{equation}
\label{eqPz0L}
\begin{split}
P_0&=I\,\Delta V,\\
P_L&=I\,\Delta V-2\frac{\rho L}{\pi R^2}I^2.
\end{split}
\end{equation}
Given that $P_L<P_0$, some of the power is entering the wires since no power is leaving $\Omega$ at infinity.

\subsection{The Power Entering the Wires}
\label{ssecPoyw}
%------------------------------------------------------------------------------

At the surfaces of the wires, the non-zero $\tau$ component of the Poynting vector in Eq.~(\ref{eqPoy2br}) indicates that power is also flowing perpendicularly to the wires' surfaces.
A positive power flow is taken as entering the wires, and with the same unit surface vector~$\hat{\vect{n}}$ as in Eq.~(\ref{eqnt}), the power density~$S_2$ at the surface of the right wire is obtained by evaluating the expression
\begin{equation}
\label{eqS2}
-\vect{S}\cdot\hat{\vect{n}}=\vect{S}\cdot\hat{\vects{\tau}}=S_\tau
\end{equation}
at the~$\tau_w$ position from Eq.~(\ref{eqtw}). Using the vertical bar to denote substitution, the power density is
\begin{equation}
\label{eqS2b}
S_2=\left(S_\tau\right)\big|_{\tau=\tau_w}
   = \left(\frac{\rho J I\tau\beta}{2\pi h_\tau\arccosh(l/R)}\right)\bigg|_{\tau=\arccosh(l/R)}.
\end{equation}

The power entering the right wire is the double integral of the power density~$S_2$ over the surface of this wire in the region~$\Omega$. With the scale factor, this integral is
\begin{equation}
\label{eqIntw1r}
P_2=\int_{0}^{2\pi}\int_{0}^{L}  \left(S_\tau h_\sigma\right)\big|_{\tau=\arccosh(l/R)}\,dz d\sigma.
\end{equation}
Since~$h_\sigma=h_\tau$, the scale factor in~$S_\tau$ cancels out with~$h_\sigma$. After evaluating the inner integral, for which the integrand has no dependency in~$z$, the power becomes
\begin{equation}
\label{eqIntw1r2}
P_2=\frac{\rho J IL}{2\pi}\int_{0}^{2\pi}\beta\big|_{\tau=\arccosh(l/R)}\,d\sigma.
\end{equation}
Any occurrence of~$\tau$ in~$\beta$ is a constant for this integral, which can alternatively be written as
\begin{equation}
\label{eqIntw1r3}
P_2=\frac{\rho J IL}{2\pi}\left(\int_{0}^{2\pi}\beta\,d\sigma\right)\bigg|_{\tau=\arccosh(l/R)}=\rho J IL.
\end{equation}
The integral in parentheses above is shown to be equal to~$2\pi$ in the appendix. Expanding the current density~$J$, the power flowing into the right wire is
\begin{equation}
\label{eqIntw1er}
P_2=\frac{\rho L }{\pi R^2}I^2\qquad (\,=P_1).
\end{equation}
A similar calculation on the left wire produces an identical result, therefore $P_1=P_2$. The above-mentioned results evoke the following quotation from Feynman \textit{et al.}~\cite{FEYNMAN1964}: ``[\ldots] \textit{the energy in a wire is flowing into the wire from the outside, rather than along the wire.}''

\subsection{Discussion}
\label{ssecDisc}
%------------------------------------------------------------------------------

The conservation of energy can now be established by verifying that the power entering~$\Omega$ is equal to the total power leaving this region. With the results in Eqs.~(\ref{eqPz0L}) and (\ref{eqIntw1er}), it is shown that
\begin{equation}
\label{eqcpr}
P_0=P_L+P_1+P_2.
\end{equation}

Calculating the surface charge density of the resistive wires can be done in a similar fashion as in Sec.~\ref{ssecDens}. For the right wire, the result is similar to Eq.~(\ref{eqSig2}), with the exception that $\Delta V$ is replaced by $\Delta V(z)$ from Eq.~(\ref{eqdVz}). Surface currents are not relevant for resistive wires in DC, and the current density inside the wires is the constant~$J$ that was used throughout this section.

A surprising generalization can be made for the wire pair: the positions of the effective thin current wires, which are located at $x=\pm l$, could actually be at $x=\pm\alpha$, where 
\begin{equation}
l-R<\alpha<l+R,
\end{equation}
with no effect on the power calculations. 
Using the symbol~$\alpha$ in place of~$l$ in Eq.~(\ref{eqvecAr0}) ultimately amounts to setting $k=a/\alpha$ in Eq.~(\ref{eqSubsk}), where it was already observed that the final results for $P_0$, $P_L$, $P_1$ or $P_2$ are independent of the value of~$k$.
The bounds for~$\alpha$ confine the thin current wires to the interiors of the thick wires both for physical reasons and to avoid any singularities in~$\beta$.
Situations in which~$\alpha$ is different than~$l$ (resistive conductors) or~$a$ (perfect conductors) have no apparent physical interpretation; however, they reveal an unexpected mathematical property of the wire pair.

%------------------------------------------------------------------------------
\section{Conclusion}
\label{sec_concl}
%------------------------------------------------------------------------------

The electrical power transported by two infinitely long and perfectly conducting wires was shown to be equal to~$I\,\Delta V$ in the electromagnetic field surrounding the wires. Currents and free charges are located on the surfaces of the wires, and because the fields are equal to zero in the perfect conductors, no power flows within them. The surface current and surface charge densities were also determined. 

When the wires have electrical resistance, a similar calculation using the Poynting vector was performed for a region separated by two planes that are perpendicular to the wires. Because of the different locations of the effective thin wires for the currents, the bipolar calculations were more involved, but showed that the power leaving the region is smaller than the power entering the region; by symbolic integration, this difference was shown to correspond to the power entering the wires from the fields. The electric field inside the wires being purely longitudinal, no power flows parallel to the wires inside of them.

In all cases, the calculations were greatly facilitated by the advantageous use of bipolar coordinates, showing once more how the choice of a proper coordinate system is an important consideration when solving electromagnetism problems. A set of~25 sample problems related to this paper is presented in the online supplementary material~\cite{SUPPMAT}.

%------------------------------------------------------------------------------
\begin{acknowledgments}
The author would like to thank: Alain H\'enault, Jean-Philippe Labb\'e, Danielle Bouthot, Genevi\`eve Savard, Kathleen Pineau, Michel Beaudin, Guillaume Roy-Fortin, Jean-S\'ebastien Closson-Duquette, and Olivier Landon-Cardinal for many discussions and comments on this topic and the paper. The author would also like to thank the anonymous reviewers and the editor for the many suggestions that helped improve the paper.

%The author has no conflicts to disclose.
\end{acknowledgments}
%------------------------------------------------------------------------------

%\appendix* % Omit the * if there's more than one appendix. 

\section*{Appendix: Definite Integral of \texorpdfstring{$\boldsymbol{\beta}$}{beta}}
%------------------------------------------------------------------------------

This appendix presents the steps used to perform the following definite integral of~$\beta$ with respect to~$\sigma$, as required for the $P_z$ and $P_2$ calculations in Sec.~\ref{sec_resist},
\begin{equation}
\label{eqIntsr}
\int_{0}^{2\pi}\beta\, d\sigma.
\end{equation}
An alternate expression for~$\beta$ is 
\begin{equation}
\label{eqBetaCplx}
\begin{split}
\beta=&\frac{k}{(k^2-1) \cos\left(\sigma+i\,\tau\right)+k^2+1}\\
        &+\frac{k}{(k^2-1) \cos\left(\sigma-i\,\tau\right)+k^2+1}.
\end{split}
\end{equation}
The above-mentioned complex form is for conciseness and to ease integration; $\beta$ is real-valued and singularity-free for any $\sigma$, $\tau$, and $k$ considered here.
Verifying the equivalences of both expressions of~$\beta$ is best done by working backward from Eq.~(\ref{eqBetaCplx}) to obtain Eq.~(\ref{eqBeta}).

Using the expression of~$\beta$ in Eq.~(\ref{eqBetaCplx}), the definite integral above is equal to~0 when evaluated in typical computer algebra systems, which is incorrect in the present context. The result of the indefinite integral of~$\beta$ is, when omitting the integration constant,
\begin{equation}
\label{eqIntar1e}
\begin{split}
F(\sigma) =&\arctan\! \left({\tan \left(\frac{\sigma}{2}+i\frac{\tau}{2}\right)}/{k}\right)\\
           &+\arctan\! \left({\tan \left(\frac{\sigma}{2}-i\frac{\tau}{2}\right)}/{k}\right).
\end{split}
\end{equation}
The calculation to be performed is now
\begin{equation}
\label{eqIntar1f}
F(2\pi)-F(0).
\end{equation}
At $\sigma=2\pi$, the $\tan()$ functions' arguments in Eq.~(\ref{eqIntar1e}) are in their second branches (with a real component of~$\pi$). The results of each $\arctan()$ must, therefore, be offset by $+\pi$, shown as follows:
\begin{equation}
\label{eqIntar1g}
\begin{split}
F(2\pi) =&\arctan\! \left({\tan \left(\pi+i\frac{\tau}{2}\right)}/{k}\right)+\pi\\
         &+\arctan\! \left({\tan \left(\pi-i\frac{\tau}{2}\right)}/{k}\right)+\pi=2\pi.
\end{split}
\end{equation}
Since $F(0)=0$, the result of Eq.~(\ref{eqIntar1f}) is $2\pi$, and it is shown that
\begin{equation}
\label{eqIntsfr}
\int_{0}^{2\pi}\beta\, d\sigma=2\pi.
\end{equation}
The complex conjugate form appearing in the simpler version of~$\beta$ above (initially obtained using Wolfram Alpha) offers a strong hint that expressing the entire problem using complex functions allows for more compact calculations. This is shown in the solution to one of the sample problems in the online supplementary material~\cite{SUPPMAT}.

\section*{References}
%------------------------------------------------------------------------------

\newcounter{prob}
\newcommand{\newprob}{\refstepcounter{prob}\textbf{\arabic{prob}.}}
\clearpage
\onecolumngrid

\setcounter{section}{0}
\setcounter{subsection}{0}

\vspace*{4mm}
\section*{Supplementary Material -- Sample Problems}

\begin{center}
\begin{minipage}{5.5in}
\textit{\rule{0pt}{8pt}Marc Boul\'e}
\\
\rule{0pt}{15pt}This document contains sample problems related to the paper \textit{DC Power Transported by Two Infinite Parallel Wires}. The problems are of varying difficulty, and the use of Maple or Mathematica software (or similar) is recommended for certain problems. Unless otherwise indicated, all references to equations numbers and figures refer to those of the paper. 
It is assumed that Eq.~(\ref{eqCenters}) can be invoked as needed to relate~$a$, $l$ and~$R$, and that when solving problems in bipolar coordinates, the wires' surfaces are located at $\tau=\pm\tau_w$ with $\tau_w$ given in Eq.~(\ref{eqtw}).
\end{minipage}
\end{center}

\vspace*{4mm}

\twocolumngrid

\subsection{Perfectly conducting wires}
%------------------------------------------------------------------------------

\newprob\label{prbVeqp}
From Eq.~(\ref{eqV}) and the equation of a circle in Cartesian coordinates, show that the wires of radius~$R$ are electric scalar equipotentials.

\newprob\label{prbAeqp}
From Eq.~(\ref{eqvecA0}) and the equation of a circle in Cartesian coordinates, show that the wires of radius~$R$ are magnetic vector equipotentials.

\newprob\label{prbCosvsCos}
Prove Eq.~(\ref{eqsigtheta}) relating $\cos\sigma$ to $\cos\theta$ for the right wire. Assume that~$\theta=\pi$ is the leftmost point of the right wire's cross section, and~$\theta=0$ (or~$2\pi$) is the rightmost point of the right wire's cross section. 

\newprob\label{prbEBcart}
Using Eqs.~(\ref{eqvecA0}) to~(\ref{eqBa}), calculate the vector expressions in Cartesian coordinates of the electric and magnetic fields outside of the wires.

\newprob\label{prbEBlaws}
Using Gauss' law and Amp\`ere's law for a single thin wire, and then the superposition principle for the thin pair of wires located at $x=\pm a$, calculate the vector expressions in Cartesian coordinates of the electric and magnetic fields outside of the wires. Assume that the line charge polarities and currents are those of Fig.~\ref{fig2d}. 

\newprob\label{prbDvpr}
Use Eq.~(\ref{eqV0}) to calculate the potential difference~$\Delta V$ between the wires as a function of~$\uplambda$, and using this result, show that the electric field obtained in Problem~\ref{prbEBlaws} is equivalent to the one obtained in Problem~\ref{prbEBcart}.

\newprob\label{prbEBw}
From the expressions obtained in Problem~\ref{prbEBcart} (or~\ref{prbEBlaws}), show that the electric and magnetic field vectors are: i) perpendicular to each other everywhere outside of the wires; ii) respectively perpendicular and parallel to the surfaces of the wires just outside of these surfaces.

\newprob\label{prbDivE}
Show that the divergence of the electric field is equal to zero in the space outside of the wires, thus confirming that no free charges exist there. Solve the problem both in Cartesian coordinates, using the electric field obtained in Problem~\ref{prbEBcart} (or~\ref{prbEBlaws}), and in bipolar coordinates, using the field in Eq.~(\ref{eqEb}) and the divergence operator:
\begin{equation*}
\nabla\cdot\vect{F} = \frac{1}{h_\sigma h_\tau h_z}\left(
\frac{\partial h_\tau h_z F_\sigma}{\partial\sigma}+
\frac{\partial h_\sigma h_z F_\tau}{\partial\tau}+
\frac{\partial h_\sigma h_\tau F_z}{\partial z}\right).
\end{equation*}

\newprob\label{prbLapE}
Show that the Laplacian of the electric potential is equal to zero in the space outside of the wires, thus confirming that no free charges exist there. Solve the problem both in Cartesian coordinates, using the electric potential in Eq.~(\ref{eqV}), and in bipolar coordinates, using the potential in Eq.~(\ref{eqVb}) and the Laplacian operator:
\begin{equation*}
\nabla^2F = \frac{1}{h_\sigma h_\tau h_z}\!\left(
\frac{\partial}{\partial\sigma}\!\left(\frac{h_\tau h_z}{h_\sigma}\frac{\partial F}{\partial\sigma}\right)\!+\!
\frac{\partial}{\partial\tau}\!\left(\frac{h_\sigma h_z}{h_\tau}\frac{\partial F}{\partial\tau}\right)\!+\!
\frac{\partial}{\partial z}\!\left(\frac{h_\sigma h_\tau}{h_z}\frac{\partial F}{\partial z}\right)\!
\right).
\end{equation*}

\newprob\label{prbDvIntE}
From the electric field obtained in Problem~\ref{prbEBcart}, confirm that the potential difference between the wires is indeed equal to the symbol~$\Delta V$, by integrating the electric field along the~$x$ axis, which corresponds to the following line integral:
\begin{equation*}
V_2-V_1=-\int_{-l+R}^{l-R}\left(E_x\right)\lvert_{y=0}\; dx.
\end{equation*}

\newprob\label{prbSzcart}
From the electric and magnetic fields obtained in Problem~\ref{prbEBcart}, show that the $z$~component of the Poynting vector in Cartesian coordinates is
\begin{equation*}
\label{eqPoyC}
S_z=\frac{I\Delta V a^2}{\pi\arccosh(l/R)\left((x-a)^2+y^2\right)\left((x+a)^2+y^2\right)}.
\end{equation*}

\newprob\label{prbRSpec}
Using~$S_z$ from Problem~\ref{prbSzcart} or the bipolar Poynting vector in Eq.~(\ref{eqPoy2b}), calculate the ratio of the power density near the inner portion of the right wire at $x=l-R$ and the outer portion of the wire at $x=l+R$, and show that in the limiting behavior $l\gg R$, this ratio is equal to~1.

\newprob\label{prbHalf}
Using the~$S_z$ expression from Problem~\ref{prbSzcart}, show that \emph{half} of the power flows within an infinite vertical rectangle tangent to the interior extremities of the wires' cross sections, as shown in Fig.~\ref{figRes} below, irrespective of the dimensions~$l$ and~$R$.
\begin{figure}[b]
\centerline{
	\begin{circuitikz}
	\draw[white, ->] (-3.9,0) -- (3.9,0); %node[anchor=west] {$x$};
	\draw[white, ->] (0,-1.5) -- (0,1.9); %node[anchor=south] {$y$};
	% coords
	\coordinate (LL) at (-1.85,0);
	\coordinate (LR) at (1.85,0);
	% big wires
	\draw[fill=lightgray] (LL) circle (0.7);
	\draw[fill=lightgray] (LR) circle (0.7);
	% vert lines
	\draw[dashed,thick] (1.15,-1.5) -- (1.15,1.86);%1.15 is 1.85-0.7
	\draw[dashed,thick] (-1.15,-1.5) -- (-1.15,1.86);%1.15 is 1.85-0.7
	% text
	\draw (0,1.2) node {\scalebox{1.33}{$\frac{I\,\Delta V}{2}$}};
	% R
	\draw (LL) -- +(-135:0.67) node[darkgray, anchor=north east] {$R$};
	\draw (LR) -- +(-45:0.67) node[darkgray, anchor=north west] {$R$};
	% l markers
	\draw[darkgray, <->] (-1.85,-0.9) -- (1.85,-0.9);
	\draw[darkgray] (0,-1.1) node {$2l$};
	\draw[darkgray, densely dashed] (-1.85,-0.05) -- (-1.85,-1.1);
	\draw[darkgray, densely dashed] (1.85,-0.05) -- (1.85,-1.1);
	\end{circuitikz}
}
\caption{Problem~\ref{prbHalf}.\label{figRes}}
\end{figure}
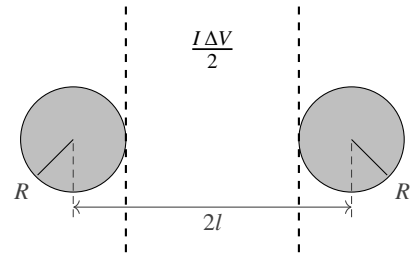

\newpage
\newprob\label{prbDV}
Using Eqs.~(\ref{eqV0}) and~(\ref{eqV}), or an intermediate result in the solution to Problem~\ref{prbDvpr}, calculate the line charge density~$\uplambda$ as a function of the potential difference~$\Delta V$. With $C^\prime=\uplambda/\Delta V$, show that the per unit length capacitance~$C^\prime$ of the wire pair is
\begin{equation*}
\label{eqC0}
C^\prime=\frac{\pi\epsilon_0}{\arccosh({l}/{R})}.
\end{equation*}

\newprob\label{prbDV2}
Repeat Problem~\ref{prbDV} by instead calculating the line charge density from
 the surface charge density~$\sigc$ using: i) the polar form in Eq.~(\ref{eqSig2}); ii) the bipolar form in Eq.~(\ref{eqSig}). Hint: integrate~$\sigc$ around the cross section of the right wire.

\newprob\label{prbWe}
Calculate the energy stored in the electric field in the space surrounding the wires for a length~$\ell$ parallel to the wires, by using the following volume integral in bipolar coordinates:
\begin{equation*}
\label{eqWEsq}
W_E=\int_0^\ell\int_0^{2\pi}\int_{-\tau_w}^{\tau_w}{\frac{1}{2}\epsilon_0\|\vect{E}\|^2}\,h_\tau h_\sigma h_z d\tau d\sigma dz,
\end{equation*}
where $\vect{E}$ is given in Eq.~(\ref{eqEb}) and the scale factors are given in Sec.~\ref{secObp}. Then show that the per unit length capacitance~$C^\prime$ in Problem~\ref{prbDV} can also be obtained from 
\begin{equation*}
\label{eqWC}
W_E=\frac{1}{2}C\Delta V^2,\quad\textrm{with~}C=C^\prime\ell.
\end{equation*}

\newprob\label{prbDA}
Calculate the difference of magnetic potential of the wires, $A_2-A_1$, labeled~$\Delta A$, and using $L^\prime=\Delta A/I$, show that the per unit length inductance~$L^\prime$ of the wire pair is 
\begin{equation*}
\label{eqL0}
L^\prime=\frac{\mu_0}{\pi}\arccosh(l/R).
\end{equation*}

\newprob\label{prbWb}
Calculate the energy stored in the magnetic field in the space surrounding the wires for a length~$\ell$ parallel to the wires, by using the following volume integral in bipolar coordinates:
\begin{equation*}
\label{eqWBsq}
W_B=\int_0^\ell\int_0^{2\pi}\int_{-\tau_w}^{\tau_w}{\frac{1}{2\mu_0}\|\vect{B}\|^2}\,h_\tau h_\sigma h_z d\tau d\sigma dz,
\end{equation*}
where $\vect{B}$ is given in Eq.~(\ref{eqBb}) and the scale factors are given in Sec.~\ref{secObp}. Then show that the per unit length inductance~$L^\prime$ in Problem~\ref{prbDA} can also be obtained from 
\begin{equation*}
\label{eqWL}
W_B=\frac{1}{2}LI^2,\quad\textrm{with~}L=L^\prime\ell.
\end{equation*}

\subsection{Resistive Wires}
%------------------------------------------------------------------------------

\newprob\label{prbABprf}
With a uniform current density~$J$ in the wires, use Amp\`ere's law for a single thin wire (and then the superposition principle) to obtain the~$\vect{B}$ field outside of the resistive wire pair in Cartesian coordinates. Show that this result is identical to the curl of~$\vect{A}$, where~$\vect{A}$ is given in Eq.~(\ref{eqvecAr0}).

\newprob\label{prbS2sig}
Using the $S_\tau$ component of the bipolar Poynting vector in Eq.~(\ref{eqPoy2br}), show that the power density~$S_2$ entering the surface of the right wire is, when expressed using the polar angle~$\theta$,
\begin{equation*}
S_2(\theta)=\frac{\left(2 l+R \cos\theta\right)\rho J I l}{\left(R^{2} + 4 l^{2} + 4 l R\cos\theta \right)\pi R}.
\end{equation*}
Assume that~$\theta=\pi$ is the leftmost point of the right wire's cross section, and~$\theta=0$ (or~$2\pi$) is the rightmost point of the right wire's cross section. Hint: see Problem~\ref{prbCosvsCos} for a useful result.

\newprob\label{prbP2alt}
Using the expression for $S_2$ in Problem~\ref{prbS2sig}, show that the power entering the right wire in the region~$\Omega$ (see Fig.~\ref{figResistance}) is equal to $\rho J I L$. Hint: calculate the double integral of the power density over the surface of this wire.

\newprob\label{prbDVr}
Perform the calculation of~$\sigma_c$ for the restive wires as indicated in Sec.~\ref{ssecDisc}. Using~$\Delta V(z)$ from Eq.~(\ref{eqdVz}) and a similar procedure as in Problem~\ref{prbDV2}, calculate the line charge density~$\uplambda(z)$ of the resistive wires. With this result, and using $C^\prime=\uplambda(z)/\Delta V(z)$, show that the per unit length capacitance of the resistive wire pair is identical to the one given in Problem~\ref{prbDV} for perfect conductors.

\newprob\label{prbRSres}
Using the $S_z$ component of the bipolar Poynting vector in Eq.~(\ref{eqPoy2br}), calculate the ratio of the longitudinal power density near the inner portion of the right wire at $x=l-R$ and the outer portion of the wire at $x=l+R$, and show that in the limiting behavior $l\gg R$, this ratio is equal to~1.

\newprob\label{prbRSres2}
Repeat Problem~\ref{prbRSres} for the transverse power entering the right wire instead of the longitudinal power flowing near the surface of the wire.

\newprob\label{prbComplex}*
Perform the calculation of the longitudinal power~$P_z$ leading to Eq.~(\ref{eqPzr}) using techniques of complex analysis, based on the complex variables
\begin{equation*}
Z=x+iy,\qquad W=\sigma+i\tau,
\end{equation*}
and the single complex transformation
\begin{equation*}
%\label{eqtz}
Z=T(W)=ia\cot\left(\frac{W}{2}\right),
\end{equation*}
along with the scale factors
\begin{equation*}
h_\sigma=h_\tau=\left|\frac{d\,T(W)}{dW}\right|=\frac{a}{|\cos(W)-1|}.
\end{equation*}
The symbol~$Z$ is a complex variable of~$x$ and~$y$, and should not be confused with the Cartesian~$z$ axis. Since complex analysis and the notations typically employed are not well suited for three dimensional calculations, use complex notation to combine both bipolar variables as much as possible, while reverting to the usual non-complex definitions of the gradient and curl operators.

\newcommand{\solprob}[1]{\subsection*{{Problem~\ref{#1}}}}
\clearpage
\onecolumngrid

\setcounter{figure}{0}
\setcounter{section}{0}
\setcounter{subsection}{0}

\vspace*{4mm}
\section*{Solutions to Sample Problems}
\vspace*{4mm}

\twocolumngrid

%------------------------------------------------------------------------------

\solprob{prbVeqp}

By substituting $y^2=R^2-(x-l)^2$ (for the right wire) and $a=\sqrt{l^2-R^2}$ into Eq.~(\ref{eqV}), the electric potential of the right wire becomes
\begin{equation*}
V_2=\frac{\Delta V}{4\arccosh(l/R)}\ln\left(\frac{l+\sqrt{l^{2}-R^{2}}}{l-\sqrt{l^{2}-R^{2}}}\right).
\end{equation*}
Since $V_2$ now contains only constants, the right wire is an equipotential. Repeating the same procedure with the substitution $y^2=R^2-(x+l)^2$ for the left wire produces the expression
\begin{equation*}
V_1=\frac{\Delta V}{4\arccosh(l/R)}\ln\left(\frac{l-\sqrt{l^{2}-R^{2}}}{l+\sqrt{l^{2}-R^{2}}}\right)=-V_2,
\end{equation*}
thus proving that the wires are both electric equipotentials.

\solprob{prbAeqp}

By substituting $y^2=R^2-(x-l)^2$ (for the right wire) and $a=\sqrt{l^2-R^2}$ into Eq.~(\ref{eqvecA0}), the magnetic vector potential of the right wire becomes
\begin{equation*}
\vect{A}_2=\frac{-\mu_0 I}{2\pi}\ln\left(\frac{\sqrt{l-\sqrt{l^{2}-R^{2}}}}{\sqrt{l+\sqrt{l^{2}-R^{2}}}}\right).
\end{equation*}
Since $\vect{A}_2$ now contains only constants, the right wire is an equipotential. Repeating the same procedure with the substitution $y^2=R^2-(x+l)^2$ for the left wire produces the expression equivalent to $-\vect{A}_2$, thus proving that the wires are both magnetic vector equipotentials.

\solprob{prbCosvsCos}

The~$x$ coordinate transformation in Eq.~(\ref{eqtx}) is evaluated at the~$\tau_w$ position of the wire given in Eq.~(\ref{eqtw}). After simplification, the~$x$ values of the points on the wire's surface are
\begin{equation*}
x=\frac{a\sqrt{l^2-R^2}}{l-R\cos\sigma},
\end{equation*}
and when using~Eq.(\ref{eqCenters}) to rewrite $a$, they are
\begin{equation*}
x=\frac{l^2-R^2}{l-R\cos\sigma}.
\end{equation*}
These~$x$ values are related to the polar angle~$\theta$ using
\begin{equation*}
x=l+R\cos\theta.
\end{equation*}
Equating the right sides of both expressions of~$x$ just above, and solving for $\cos\sigma$ produces
\begin{equation*}
\cos\sigma = \frac{R+l\cos\theta}{l + R\cos\theta}.
\end{equation*}

\solprob{prbEBcart}

The electric scalar potential~$V$ and magnetic vector potential~$\vect{A}$ from Eqs.~(\ref{eqV}) and~(\ref{eqvecA0}) respectively can be used to calculate the electric and magnetic fields directly in Cartesian coordinates using Eqs.~(\ref{eqEphi}) and~(\ref{eqBa}):
\begin{equation*}
\vect{E}=-\nabla V = 
\frac{\left((x^2-y^2-a^2)\,\hat{\vect{x}}+2xy\,\hat{\vect{y}}\right)a\Delta V}{\left((x-a)^2+y^2\right)\left((x+a)^2+y^2\right)\arccosh(l/R)},
\end{equation*}
\begin{equation*}
\vect{B}=\vects{\nabla}\times\vect{A} = 
\frac{\left(-2xy\,\hat{\vect{x}}+(x^2-y^2-a^2)\,\hat{\vect{y}}\right)\mu_0 a I}{\left((x-a)^2+y^2\right)\left((x+a)^2+y^2\right)\pi}.
\end{equation*}

\solprob{prbEBlaws}

The electric and magnetic fields produced by the wire pair can be obtained from two fundamental laws in electromagnetism, namely Gauss' law and Amp\`ere's law, which are two of Maxwell's four foundational equations. These two laws, when applied to a straight thin wire, are solved in almost any textbook dealing with electromagnetism. From these, the field magnitudes around a single thin wire are proved to be
\begin{equation}
\label{eqAeb}
E=\frac{\uplambda}{2\pi\epsilon_0 r},\quad \textrm{and}\quad B=\frac{\mu_0 I}{2\pi r},
\end{equation}
where~$r$ is the distance from the axis of the thin wire. The electric field is radial to the wire, while the magnetic field is circular around the wire, according to the right-hand rule.

Considering two thin wires parallel to the~$z$ axis, located at~$x=\pm a$, the superposition principle allows the calculation of the resultant electric and magnetic field vectors. The same charge densities and current directions as those appearing in Fig.~\ref{fig2d} are used here. The left wire is designated wire~1, and the right wire is designated wire~2. The vector positions of the thin wires are
\begin{equation*}
\vect{r}_{w1}=-a \,\hat{\vect{x}},\quad \textrm{and}\quad \vect{r}_{w2}=a \,\hat{\vect{x}}.
\end{equation*}
The vector fields are calculated at the arbitrary point
\begin{equation*}
\vect{p}=x\,\hat{\vect{x}}+y\,\hat{\vect{y}}.
\end{equation*}
The distances between this point and the wires are required. In vector form, these are
\begin{equation*}
\vect{r}_{1}=\vect{p}-\vect{r}_{w1}=(x+a) \,\hat{\vect{x}}+y\,\hat{\vect{y}},
\end{equation*}
\begin{equation*}
\vect{r}_{2}=\vect{p}-\vect{r}_{w2}=(x-a) \,\hat{\vect{x}}+y\,\hat{\vect{y}}.
\end{equation*}
From the vector distances above, the fields of wire~1, which has a negative charge density and a current in~$-\hat{\vect{z}}$, are
\begin{equation*}
\vect{E}_1=\frac{-\uplambda}{2\pi\epsilon_0\lvert\vect{r}_{1}\rvert}\frac{\vect{r}_{1}}{\lvert\vect{r}_{1}\rvert}
= \frac{-\uplambda}{2\pi\epsilon_0\left((x+a)^2+y^2\right)}\left((x+a)\,\hat{\vect{x}} + y\,\hat{\vect{y}}\right),
\end{equation*}
\begin{equation*}
\vect{B}_1=\frac{\mu_0 I}{2\pi\lvert\vect{r}_{1}\rvert}\frac{(-\hat{\vect{z}})\times\vect{r}_{1}}{\lvert\vect{r}_{1}\rvert}
= \frac{\mu_0 I}{2\pi\left((x+a)^2+y^2\right)}\left(y\,\hat{\vect{x}} + -(x+a)\,\hat{\vect{y}}\right).
\end{equation*}
Similarly, the fields of wire~2, which has a positive charge density and a current in~$\hat{\vect{z}}$, are
\begin{equation*}
\vect{E}_2 =\frac{\uplambda}{2\pi\epsilon_0\lvert\vect{r}_{2}\rvert}\frac{\vect{r}_{2}}{\lvert\vect{r}_{2}\rvert}
= \frac{\uplambda}{2\pi\epsilon_0\left((x-a)^2+y^2\right)}\left((x-a)\,\hat{\vect{x}} + y\,\hat{\vect{y}}\right),
\end{equation*}
\begin{equation*}
\vect{B}_2=\frac{\mu_0 I}{2\pi\lvert\vect{r}_{2}\rvert}\frac{\hat{\vect{z}}\times\vect{r}_{2}}{\lvert\vect{r}_{2}\rvert}
= \frac{\mu_0 I}{2\pi\left((x-a)^2+y^2\right)}\left(-y\,\hat{\vect{x}} + (x-a)\,\hat{\vect{y}}\right).
\end{equation*}
The resultant fields are the vector sums of the fields from each wire. When simplified, these field are
\begin{equation*}
\label{eqAEv}
\vect{E}=
\frac{\left((x^2-y^2-a^2)\,\hat{\vect{x}}+2xy\,\hat{\vect{y}}\right)a\uplambda}{\left((x-a)^2+y^2\right)\left((x+a)^2+y^2\right)\pi\epsilon_0},
\end{equation*}
\begin{equation*}
\label{eqABv}
\vect{B}= 
\frac{\left(-2xy\,\hat{\vect{x}}+(x^2-y^2-a^2)\,\hat{\vect{y}}\right)\mu_0 a I }{\left((x-a)^2+y^2\right)\left((x+a)^2+y^2\right)\pi}.
\end{equation*}

\solprob{prbDvpr}

Evaluating the potential
\begin{equation*}
V=\frac{-\uplambda}{2\pi\epsilon_0}\ln\left(\frac{\sqrt{\left(x-a\right)^2+y^2}}{\sqrt{\left(x+a\right)^2+y^2}}\right),
\end{equation*}
at position $(-l+R,0)$ and $(l-R,0)$, yields the potentials of the wires $V_1$ and $V_2$ respectively:
\begin{equation*}
V_1=\frac{-\uplambda}{2\pi\epsilon_0}\ln\left(\frac{a+(l-R)}{a-(l-R)}\right),
\end{equation*}
\begin{equation*}
V_2=\frac{\uplambda}{2\pi\epsilon_0}\ln\left(\frac{a+(l-R)}{a-(l-R)}\right).
\end{equation*}
The potential difference between the wires is
\begin{equation*}
\Delta V = V_2-V_1=\frac{\uplambda}{\pi\epsilon_0}\ln\left(\frac{a+(l-R)}{a-(l-R)}\right),
\end{equation*}
which, using Eq.~(\ref{eqCenters}) and simplifying, can be written as
\begin{equation*}
\Delta V=\frac{\uplambda}{\pi\epsilon_0}\ln\left(\frac{l+\sqrt{l^2-R^2}}{R}\right),
\end{equation*}
or equivalently,
\begin{equation*}
\Delta V=\frac{\uplambda}{\pi\epsilon_0}\ln\left(\frac{l}{R}+\sqrt{\left(\frac{l}{R}\right)^2-1}\right).
\end{equation*}
Invoking the trigonometric identity
\begin{equation*}
\arccosh(u)=\ln\left(u+\sqrt{u^2-1}\right),
\end{equation*}
the potential difference between the wires becomes
\begin{equation*}
\Delta V=\frac{\uplambda}{\pi\epsilon_0}\arccosh\!\left(\frac{l}{R}\right).
\end{equation*}
Solving for the line charge density~$\uplambda$,
\begin{equation*}
\uplambda=\frac{\pi\epsilon_0\Delta V}{\arccosh(l/R)},
\end{equation*}
and substituting this in the expression of the electric field~$\vect{E}$ obtained in Problem~\ref{prbEBlaws} results in an expression that is identical to the~$\vect{E}$ field obtained in Problem~\ref{prbEBcart}.

\solprob{prbEBw}

i) The~$\vect{E}$ and~$\vect{B}$ vectors are perpendicular to each other outside of the wires if their dot product is equal to zero. Using results Problem~\ref{prbEBcart} or~\ref{prbEBlaws}, it is shown that
\begin{equation*}
\vect{E}\cdot\vect{B}=0.
\end{equation*}

ii) To show that the~$\vect{E}$ field is perpendicular to the surface of the right wire, a normal vector can be established using a vector difference of the positions of the points on the wire and the center of the wire. For the left and right wires, these vector radii are:
\begin{equation*}
\vect{R_1}=(x+l)\,\hat{\vect{x}}+\left(\sqrt{R^2-(x+l)^2}\right)\,\hat{\vect{y}},
\end{equation*}
\begin{equation*}
\vect{R_2}=(x-l)\,\hat{\vect{x}}+\left(\sqrt{R^2-(x-l)^2}\right)\,\hat{\vect{y}}.
\end{equation*}
Substituting the~$y$ value of the cross section of the left wire
\begin{equation*}
y=\sqrt{R^2-(x+l)^2}
\end{equation*}
into the $\vect{E}$ field, along with the value of~$a$ from Eq.~(\ref{eqCenters}), the electric field at the surface of the left wire is
\begin{equation*}
\vect{E}_{w1}=\frac{\Delta V \sqrt{l^2-R^2}}{2xR^2\arccosh(l/R)}\left((x+l)\,\hat{\vect{x}}+\sqrt{R^2-(x+l)^2}\,\hat{\vect{y}}\right).
\end{equation*}
It can then be verified that 
\begin{equation*}
\vect{E}_{w1}\times\vect{R}_1=\vect{0},
\end{equation*}
proving that the electric field is parallel to the vector radius of the left wire, thus perpendicular to the wire's surface. 
Similarly, substituting the $y$ value of the cross section of the right wire into the $\vect{E}$ field, the electric field at the surface of the right wire is
\begin{equation*}
\vect{E}_{w2}=\frac{\Delta V \sqrt{l^2-R^2}}{2xR^2\arccosh(l/R)}\left((x-l)\,\hat{\vect{x}}+\sqrt{R^2-(x-l)^2}\,\hat{\vect{y}}\right).
\end{equation*}
It can then be verified that 
\begin{equation*}
\vect{E}_{w2}\times\vect{R}_2=\vect{0},
\end{equation*}
proving that the electric field is parallel to the vector radius of the right wire, thus perpendicular to the wire's surface. In both proofs above, the top portion of the wires was used; however, the same results are obtained using the bottom portion of the wires. For example, for the left wire, the~$y$ value would be
\begin{equation*}
y=-\sqrt{R^2-(x+l)^2},
\end{equation*}
which would be substituted in~$\vect{E}$, and the vector radius would be
\begin{equation*}
\vect{R_1}=(x+l)\,\hat{\vect{x}}-\left(\sqrt{R^2-(x+l)^2}\right)\,\hat{\vect{y}}.
\end{equation*}
To show that the~$\vect{B}$ field is tangent to the surfaces of the wires, the procedure is very similar, with the exception that the $\vect{B}$ field is used, and the condition that must be verified is the dot product. For both wires, it can be shown that
\begin{equation*}
\vect{B}_{w1}\cdot\vect{R}_1=0,\qquad\vect{B}_{w2}\cdot\vect{R}_2=0.
\end{equation*}

\solprob{prbDivE}

For the Cartesian solution, with 
\begin{equation*}
\vect{E}= 
\frac{\left((x^2-y^2-a^2)\,\hat{\vect{x}}+2xy\,\hat{\vect{y}}\right)a\Delta V}{\left((x-a)^2+y^2\right)\left((x+a)^2+y^2\right)\arccosh\frac{l}{R}},
\end{equation*}
it can be shown that $\nabla\cdot\vect{E}=0$ using the usual divergence operator. In bipolar coordinates, the more general divergence operator given in the problem's statement must be used. Its expression can be simplified given that $h_z=1$ and that the electric field in Eq.~(\ref{eqEb}) only has a single $\hat{\vects{\tau}}$ component:
\begin{equation*}
\nabla\cdot\vect{E} = \frac{1}{h_\sigma h_\tau}\left(
\frac{\partial h_\sigma E_\tau}{\partial\tau}\right),
\end{equation*}
which can be shown to also be equal to~0.

\solprob{prbLapE}

The solution follows naturally using the usual definition of the Laplacian operator, as it did in Problem~\ref{prbDivE} for the divergence. The same can be said for the solution in bipolar coordinates, where the more general Laplacian operator given in the problem's statement must be used. Its expression can be simplified given that $h_z=1$, $h_\sigma=h_\tau$, and that the electric potential in Eq.~(\ref{eqVb}) depends only on $\tau$:
\begin{equation*}
\nabla^2 V = \frac{1}{h_\sigma h_\tau}
\frac{\partial}{\partial\tau}\!\left(\frac{\partial V}{\partial\tau}\right),
\end{equation*}
which can be shown to also be equal to~0.
Alternatively for both solutions, since the Laplacian operator is defined as:
\begin{equation*}
\nabla^2 V = \nabla\cdot(\nabla V),
\end{equation*}
and since, $\vect{E}=-\nabla V$ (in either coordinate system), the problem can be expressed using divergence as
\begin{equation*}
\nabla^2 V = -\nabla\cdot\vect{E},
\end{equation*}
where is was shown in Problem~\ref{prbDivE} that the divergence of~$\vect{E}$ is equal to~0, thereby also proving that $\nabla^2 V=0$ in both coordinate systems.

\solprob{prbDvIntE}

The electric field obtained in Problem~\ref{prbEBcart} is evaluated at the position $y=0$, where, after simplification, the~$x$ component is:
\begin{equation*}
E_x(x,0)=-\frac{a\Delta V}{\left(a^2-x^2\right)\arccosh(l/R)}.
\end{equation*}
The integral to perform is
\begin{equation*}
V_2-V_1=-\int_{-l+R}^{l-R}E_x(x,0)\,dx.
\end{equation*}
Since $E_x(x,0)$ is an even function in~$x$, the integral reduces to
\begin{equation*}
V_2-V_1=\frac{2\Delta V}{\arccosh(l/R)}\int_0^{l-R}\frac{a}{\left(a^2-x^2\right)}\,dx.
\end{equation*}
The integrand is converted to partial fractions to facilitate integration:
\begin{equation*}
V_2-V_1=\frac{\Delta V}{\arccosh(l/R)}\int_0^{l-R}\left(\frac{1}{a+x}+\frac{1}{a-x}\right)dx.
\end{equation*}
Performing the indefinite integral yields the following primitive (omitting the integration constant):
\begin{equation*}
F(x)=\frac{\Delta V}{\arccosh(l/R)}\ln\left(\frac{a+x}{a-x}\right).
\end{equation*}
When evaluating the bounds and simplifying,
\begin{equation*}
V_2-V_1=F(l-R)-F(0)=\frac{\Delta V}{\arccosh(l/R)}\ln\left(\frac{a+(l-R)}{a-(l-R)}\right).
\end{equation*}
As developed in the solution to Problem~\ref{prbDvpr}, the $\ln()$ above, which is the result of the integral itself, can be expressed more succinctly:
\begin{equation*}
\ln\left(\frac{a+(l-R)}{a-(l-R)}\right)=\arccosh(l/R).
\end{equation*}
The final expression for the potential difference thus reduces to:
\begin{equation*}
V_2-V_1=\Delta V.
\end{equation*}

\solprob{prbSzcart}

The Poynting vector is calculated with the cross product as follows
\begin{equation*}
\vect{S}=\frac{1}{\mu_0}\left(\vect{E}\times\vect{B}\right).% Griffiths p.358
\end{equation*}
Using the $\vect{E}$ and $\vect{B}$ fields obtained in the solution to Problem~\ref{prbEBcart}, the Poynting vector becomes
\begin{equation*}
\vect{S}=\frac{I\Delta V a^2}{\pi\arccosh(l/R)\left((x-a)^2+y^2\right)\left((x+a)^2+y^2\right)}\hat{\vect{z}}.
\end{equation*}
The $z$ component of the Poynting vector in Cartesian coordinates is therefore
\begin{equation*}
S_z=\frac{I\Delta V a^2}{\pi\arccosh(l/R)\left((x-a)^2+y^2\right)\left((x+a)^2+y^2\right)}.
\end{equation*}

\solprob{prbRSpec}

The ratio of power densities, labeled $f$ below, is first calculated using the expression of $S_z=S_z(x,y)$ given in Problem~\ref{prbSzcart}. The ratio required is obtain by the calculation
\begin{equation*}
f=\frac{S_z(l-R,0)}{S_z(l+R,0)}\bigg|_{a=\sqrt{l^2-R^2}},
\end{equation*}
which evaluates to, when simplified,
\begin{equation*}
f=\frac{(l+R)^2}{(l-R)^2}.
\end{equation*}
When $l\gg R$, $f$ is equal to~1 and it is thus shown that the same amount of power flows locally near the inner portion of the surface and the outer portion. 
The same expression of $f$ can also be obtained by using bipolar form $S_z=S_z(\sigma,\tau)$ from Eq.~(\ref{eqPoy2b}). In this case, the ratio is calculated as
\begin{equation*}
f=\frac{S_z(\pi,\tau_w)}{S_z(0,\tau_w)}\bigg|_{a=\sqrt{l^2-R^2},~\tau_w=\arccosh(l/R)},
\end{equation*}
which also evaluates to, when simplified,
\begin{equation*}
f=\frac{(l+R)^2}{(l-R)^2}.
\end{equation*}

\solprob{prbHalf}

The power flowing perpendicular to the $z$ plane is the surface integral of the Poynting vector in the space outside of the wires, which for the present problem consists of an infinite vertical rectangle tangent to the interior extremities of the wires' cross sections. 
Considering first only the region in the first quadrant, the bounds of the double integral are
\begin{equation*}
P=\int_0^{l - R}\int_0^\infty{S_z\,dy dx}.
\end{equation*}
Making use of the~$z$ component of the Poynting vector given in Problem~\ref{prbSzcart}, the inner integral is first evaluated. With the constraint $0<x<a$, the result is
\begin{equation*}
\int_0^\infty S_z\,dy =\frac{I\Delta V a}{4\arccosh(l/R)\left(a^2-x^2\right)}\textrm{\quad when~}0<x<a.
\end{equation*}
This result is converted to partial fractions
\begin{equation*}
\frac{I\Delta V}{8\arccosh(l/R)}\left(\frac{1}{a+x}+\frac{1}{a-x}\right),
\end{equation*}
to facilitate the calculation of the outer integral (according to~$x$). The power then becomes
\begin{equation*}
P =\frac{I\Delta V}{8\arccosh(l/R)}\int_0^{l-R}\left(\frac{1}{a+x}+\frac{1}{a-x}\right)dx.
\end{equation*}
The integral above was shown to reduce to $\arccosh(l/R)$ in Problem~\ref{prbDvIntE}, thus revealing that
\begin{equation*}
P =\frac{I\,\Delta V}{8}.
\end{equation*}
Extending this to all four quadrants indicates that \emph{one half} of the power is transported within an infinite vertical rectangle tangent to the interior extremities of the wires' cross sections, irrespective of the dimensions~$l$ and~$R$.

\solprob{prbDV}

In Problem~\ref{prbDvpr} is was shown that the line charge density~$\uplambda$ relates to the potential difference between the wires as
\begin{equation*}
\uplambda=\frac{\pi\epsilon_0\Delta V}{\arccosh(l/R)}.
\end{equation*}
The above expression can alternatively be obtained by modifying Eq.~(\ref{eqV0}) as follows
\begin{equation*}
V=\frac{\uplambda}{4\pi\epsilon_0}\ln\left(\frac{{\left(x+a\right)^2+y^2}}{{\left(x-a\right)^2+y^2}}\right),% derived from (Haus 4.6.eq18, p.151).
\end{equation*}
and then by a visual comparison with Eq.~(\ref{eqV}):
\begin{equation*}
V=\frac{\Delta V}{4\arccosh(l/R)}\ln\left(\frac{{\left(x+a\right)^2+y^2}}{{\left(x-a\right)^2+y^2}}\right),
\end{equation*}
the expression of~$\uplambda$ further above is also confirmed. Once established, the per unit length capacitance~$C^\prime$ of the perfectly conducting wire pair is
\begin{equation*}
C^\prime=\frac{\uplambda}{\Delta V}=\frac{\pi\epsilon_0}{\arccosh({l}/{R})}.
\end{equation*}

\solprob{prbDV2}

i) Integrating the polar form of the surface charge density~$\sigma_c$ of the right wire in Eq.~(\ref{eqSig2}) around the cross section of that wire produces the line charge density~$\uplambda$, as follows
\begin{equation*}
\uplambda=\int_0^{2\pi}\sigc R\,d\theta=\int_0^{2\pi}\frac{\epsilon_0\Delta V\sqrt{l^2-R^2}}{2\arccosh(l/R)\,(l+R\cos\theta)}\,d\theta.
\end{equation*}
Since computer algebra systems typically evaluate the integral above to 0 (which is incorrect in the present context), performing the indefinite integral and applying the bounds manually is preferred.
The indefinite integral yields, omitting the integration constant,
\begin{equation*}
F(\theta)=\frac{\epsilon_0\Delta V}{\arccosh(l/R)}\arctan\!\left(\sqrt{\frac{l-R}{l+R}}\tan\left(\frac{\theta}{2}\right)\right).
\end{equation*}
The calculation to be performed is now
\begin{equation*}
\uplambda=F(2\pi)-F(0).
\end{equation*}
At $\theta=2\pi$, the $\tan()$ function's argument is in its second branch (with an argument of~$\pi$). The result of the $\arctan()$ must therefore be offset by $+\pi$, as shown below
\begin{equation*}
F(2\pi)=\frac{\epsilon_0\Delta V}{\arccosh(l/R)}\left(\arctan\!\left(\sqrt{\frac{l-R}{l+R}}\tan\left(\pi\right)\right)+\pi\right),
\end{equation*}
which evaluates to
\begin{equation*}
F(2\pi)=\frac{\pi\epsilon_0\Delta V}{\arccosh(l/R)}.
\end{equation*}
Since $F(0)=0$, it is proven that the line charge density~$\uplambda$ is
\begin{equation*}
\uplambda=\frac{\pi\epsilon_0\Delta V}{\arccosh(l/R)}.
\end{equation*}
From this result, the expected capacitance is obtained
\begin{equation*}
C^\prime=\frac{\uplambda}{\Delta V}=\frac{\pi\epsilon_0}{\arccosh({l}/{R})}.
\end{equation*}

ii) Before performing the integral, the scale factor $h_\sigma$ of Eq.~(\ref{eqscst}), must be evaluated at the~$\tau$ coordinate of the right wire~($\tau_w$). The required scale factor, called simply~$h$, is
\begin{equation*}
h=h_\sigma\big|_{\tau=\tau_w} = \frac{a}{\cosh\tau-\cos\sigma}\Big|_{\tau=\arccosh(l/R)},
\end{equation*}
which becomes
\begin{equation*}
h=\frac{a}{l/R-\cos\sigma}.
\end{equation*}
Integrating the bipolar form of the surface charge density~$\sigma_c$ of the right wire in Eq.~(\ref{eqSig}) around the cross section of that wire produces the line charge density~$\uplambda$, as follows
\begin{equation*}
\uplambda=\int_0^{2\pi}\sigc h\,d\sigma=\int_0^{2\pi}\frac{\epsilon_0 \Delta V a}{2\arccosh(l/R)\sqrt{l^2-R^2}}\,d\sigma.
\end{equation*}
Simplifying, given that $a=\sqrt{l^2-R^2}$, and moving the constants out of the integral, the integral itself evaluates to $2\pi$, thus proving also that the line charge density is
\begin{equation*}
\uplambda=\frac{\pi\epsilon_0\Delta V}{\arccosh(l/R)},
\end{equation*}
from which the same expected capacitance is also obtained.

\solprob{prbWe}

The squared magnitude of the electric field vector~$\vect{E}$ is required for the energy integral. With the field in Eq.~(\ref{eqEb}), the magnitude is
\begin{equation*}
\|\vect{E}\|=\frac{(\cosh\tau-\cos\sigma)\Delta V}{2a\arccosh(l/R)},
\end{equation*}
and the squared magnitude is
\begin{equation*}
\|\vect{E}\|^2=\frac{(\cosh\tau-\cos\sigma)^2}{a^2}\frac{\Delta V^2}{4(\arccosh(l/R))^2}.
\end{equation*}
The volume integral to calculate is
\begin{equation*}
W_E=\int_0^\ell\int_0^{2\pi}\int_{-\tau_w}^{\tau_w}{\frac{1}{2}\epsilon_0\|\vect{E}\|^2}\,h_\tau h_\sigma h_z d\tau d\sigma dz.
\end{equation*}
Given that the scale factors $h_\sigma$ and $h_\tau$ from Sec.~\ref{secObp} cancel out with the left side fraction in $\|\vect{E}\|^2$, and with $h_z=1$, the integrand is much simpler
\begin{equation*}
W_E=\int_0^\ell\int_0^{2\pi}\int_{-\tau_w}^{\tau_w}\frac{\epsilon_0\Delta V^2}{8(\arccosh(l/R))^2}\, d\tau d\sigma dz.
\end{equation*}
With $\tau_w=\arccosh(l/R)$ the integral above evaluates to
\begin{equation*}
W_E=\frac{\pi\epsilon_0\Delta V^2\ell}{2\arccosh(l/R)}.
\end{equation*}
The energy in the electric field is related to the per-unit length capacitance $C^\prime$ according to
\begin{equation*}
W_E=\frac{1}{2}C^\prime\ell\Delta V^2.
\end{equation*}
From the equivalence of the two different forms of $W_E$ above, the per-unit length capacitance from Problem~\ref{prbDV} can be confirmed:
\begin{equation*}
C^\prime=\frac{\pi\epsilon_0}{\arccosh({l}/{R})}.
\end{equation*}

\solprob{prbDA}

A magnetic potential difference between the wires can be defined in a similar manner as the electric potential difference. This is possible since the magnetic vector potential~$\vect{A}$ is constant on the entire surface of each wire and is everywhere parallel to the wires. Evaluating the~$z$ component of~$\vect{A}$ in Eq.~(\ref{eqvecAb}) at the coordinate $\tau=\tau_w$ from Eq.~(\ref{eqtw}) yields the magnetic potential of the right wire
\begin{equation*}
A_2 = \frac{\mu_0 I}{2\pi}\arccosh(l/R).
\end{equation*}
Evaluating the~$z$ component of~$\vect{A}$ at $\tau=-\tau_w$, the magnetic potential of the left wire can be shown to be $A_1=-A_2$. 
The difference of magnetic potential $A_2-A_1$ of the wires, labeled~$\Delta A$, is
\begin{equation*}
\Delta A =\frac{\mu_0 I}{\pi}\arccosh(l/R).
\end{equation*}
The per-unit length inductance becomes
\begin{equation*}
L^\prime=\frac{\Delta A}{I}=\frac{\mu_0}{\pi}\arccosh(l/R).
\end{equation*}

\solprob{prbWb}

The squared magnitude of the magnetic field vector~$\vect{B}$ is required for the energy integral. With the field in Eq.~(\ref{eqBb}), the magnitude is
\begin{equation*}
\|\vect{B}\|=\frac{(\cosh\tau-\cos\sigma)\mu_0 I}{2\pi a},
\end{equation*}
and the squared magnitude is
\begin{equation*}
\|\vect{B}\|^2=\frac{(\cosh\tau-\cos\sigma)^2}{a^2}\frac{(\mu_0)^2 I^2}{4\pi^2}.
\end{equation*}
The volume integral to calculate is
\begin{equation*}
W_B=\int_0^\ell\int_0^{2\pi}\int_{-\tau_w}^{\tau_w}{\frac{1}{2\mu_0}\|\vect{B}\|^2}\,h_\tau h_\sigma h_z d\tau d\sigma dz.
\end{equation*}
Given that the scale factors $h_\sigma$ and $h_\tau$ from Sec.~\ref{secObp} cancel out with the left side fraction in $\|\vect{B}\|^2$, and with $h_z=1$, the integrand is much simpler
\begin{equation*}
W_B=\int_0^\ell\int_0^{2\pi}\int_{-\tau_w}^{\tau_w}{\frac{\mu_0I^2}{8\pi^2}}\, d\tau d\sigma dz.
\end{equation*}
With $\tau_w=\arccosh(l/R)$ the integral above evaluates to
\begin{equation*}
W_B=\frac{\mu_0I^2\ell}{2\pi}\arccosh(l/R).
\end{equation*}
The energy in the magnetic field is related to the per-unit length inductance $L^\prime$ with
\begin{equation*}
W_B=\frac{1}{2}L^\prime\ell I^2.
\end{equation*}
From the equivalence of the two different forms of $W_B$ above, the per-unit length inductance from Problem~\ref{prbDA} can be confirmed:
\begin{equation*}
L^\prime=\frac{\Delta A}{I}=\frac{\mu_0}{\pi}\arccosh(l/R).
\end{equation*}

%------------------------------------------------------------------------------

\solprob{prbABprf}

With the uniform current density~$J$ in the wires, the calculation of Amp\`ere's law is similar to that of Problem~\ref{prbEBlaws}, with the exception that the true centers of the wires must be used. For a single thick wire, the center of symmetry is located at the center of the wire's cross section. Following a similar calculation as in Problem~\ref{prbEBlaws}, the result obtained is
\begin{equation*}
\vect{B}= 
\frac{\left(-2xy\,\hat{\vect{x}}+(x^2-y^2-l^2)\,\hat{\vect{y}}\right)\mu_0 l I }{\left((x-l)^2+y^2\right)\left((x+l)^2+y^2\right)\pi}.
\end{equation*}
A comparison with the field obtained using $\vects{\nabla}\times\vect{A}$ must be done, where from Eq.~(\ref{eqvecAr0}), $\vect{A}$ is:
\begin{equation*}
\vect{A}=\frac{-\mu_0 I}{2\pi}\ln\left(\frac{\sqrt{\left(x-l\right)^2+y^2}}{\sqrt{\left(x+l\right)^2+y^2}}\right)\hat{\vect{z}}.
\end{equation*}
In Cartesian coordinates, the curl operator is
\begin{equation*}
\vects{\nabla}\times\vect{A} = \begin{vmatrix}
\hat{\vect{x}} & \hat{\vect{y}} & \hat{\vect{z}}\\ 
\rule{0pt}{13pt}\frac{\partial}{\partial x} & \frac{\partial}{\partial y} & \frac{\partial}{\partial z}\\
\rule{0pt}{12pt}A_x & A_y & A_z
\end{vmatrix},
\end{equation*}
which can be simplified since $A_x = A_y = 0$. Performing the calculation yields
\begin{equation*}
\vect{B}=\vects{\nabla}\times\vect{A}=\frac{\left(-2xy\,\hat{\vect{x}}+(x^2-y^2-l^2)\,\hat{\vect{y}}\right)\mu_0 l I }{\left((x-l)^2+y^2\right)\left((x+l)^2+y^2\right)\pi},
\end{equation*}
which is identical to the~$\vect{B}$ field obtained using Amp\`ere's law.

\solprob{prbS2sig}

The solution begins by applying the following substitutions to the~$S_\tau$ component in Eq.~(\ref{eqPoy2br}):
\begin{enumerate}
\item The substitutions for~$h_\tau$ and~$\beta$ from Eqs.~(\ref{eqscst}) and~(\ref{eqBeta}), respectively;
\item The substitution $\tau=\tau_w=\arccosh(l/R)$;
\item The substitution $k=a/l$ from Eq.~(\ref{eqSubsk});
\item The substitution $a=\sqrt{l^{2}-R^{2}}$.
\end{enumerate}
After these steps, the power density at the surface of the right wire is
\begin{equation*}
\label{eqS2ss}
S_2(\sigma)=\frac{\left(R^{2} - 2 l^{2} + l R \cos\sigma\right)\rho J I l }{\left(3 l R^{2} - 4 l^{3} + R^{3} \cos\sigma \right)\pi R }.
\end{equation*}
Using Eq.~(\ref{eqsigtheta}), which was proved in Problem~\ref{prbCosvsCos}, the result can also be expressed as
\begin{equation*}
S_2(\theta)=\frac{\left(2 l+R \cos\theta\right)\rho J I l}{\left(R^{2} + 4 l^{2} + 4 l R\cos\theta \right)\pi R}.
\end{equation*}

\solprob{prbP2alt}

The power entering the right wire is the double integral of the power density over the surface of this wire in the region~$\Omega$:
\begin{equation*}
P_2=\int_{0}^{L}\int_{0}^{2\pi}  S_2(\theta)R  \, d\theta\, dz.
\end{equation*}
Inverting the order of integration and evaluating the integral according to~$z$, $P_2$ becomes
\begin{equation*}
P_2=\int_{0}^{2\pi}  S_2(\theta)L\,R  \, d\theta.
\end{equation*}
Since some computer algebra systems evaluate the integral above to 0, performing the indefinite integral and applying the bounds manually is preferred.
The indefinite integral yields, omitting the integration constant,
\begin{equation*}
\begin{split}
F(\theta)=\frac{\rho J I L}{2\pi}\Bigg(&\arctan\!\left(\frac{\tan\left(\frac{\theta}{2}\right) \left(2 l-R\right)}{2 l+R}\right)\\
&+\arctan\!\left(\tan\left(\frac{\theta}{2}\right)\right)\Bigg).
\end{split}
\end{equation*}
The calculation to be performed is now
\begin{equation*}
P_2=F(2\pi)-F(0).
\end{equation*}
At $\theta=2\pi$, the $\tan()$ functions' arguments are in their second branches (with an argument of~$\pi$). The results of the $\arctan()$ must therefore be offset by $+\pi$, as shown below
\begin{equation*}
\begin{split}
F(2\pi)=\frac{\rho J I L}{2\pi}\Bigg(&\arctan\!\left(\frac{\tan\left(\pi\right) \left(2 l-R\right)}{2 l+R}\right)+\pi\\
&+\arctan\!\left(\tan\left(\pi\right)\right)+\pi\Bigg)=\rho J I L.
\end{split}
\end{equation*}
Since $F(0)=0$, it is proven that the power entering the right wire in the region~$\Omega$ is equal to $\rho J I L$.

\solprob{prbDVr}

The procedure for obtaining the surface charge density~$\sigc$ of the right resistive wire is described in Sec.~\ref{ssecDisc}. The density is similar to that of the perfectly conducting wire, with the exception that~$\Delta V$ is replaced by~$\Delta V(z)$ from Eq.~(\ref{eqdVz}), which is
\begin{equation*}
\Delta V(z) = \Delta V - 2\rho J z.
\end{equation*}
When this is inserted into~$\sigc$ from Eq.~(\ref{eqSig2}), in place of~$\Delta V$, the surface charge density of the right resistive wire is
\begin{equation*}
\sigc=\frac{\epsilon_0(\Delta V - 2\rho J z)\sqrt{l^2-R^2}}{2R\arccosh(l/R)\,(l+R\cos\theta)}.
\end{equation*}
Integrating the polar form above around the cross section of that wire produces the line charge density~$\uplambda(z)$, as follows
\begin{equation*}
\uplambda(z)=\int_0^{2\pi}\sigc R\,d\theta=\int_0^{2\pi}\frac{\epsilon_0(\Delta V - 2\rho J z)\sqrt{l^2-R^2}}{2\arccosh(l/R)\,(l+R\cos\theta)}\,d\theta.
\end{equation*}
Following the same procedure as in Problem~\ref{prbDV2}~i), the charge density obtained is 
\begin{equation*}
\uplambda(z)=\frac{\pi\epsilon_0(\Delta V - 2\rho J z)}{\arccosh(l/R)}.
\end{equation*}
The per-unit length capacitance of the resistive wire pair can be calculated: 
\begin{equation*}
C^\prime=\frac{\uplambda(z)}{\Delta V(z)}=\frac{\pi\epsilon_0}{\arccosh(l/R)},
\end{equation*}
which is the same capacitance as for the perfectly conducting wire pair.

\solprob{prbRSres}

The ratio of power densities, labeled $f$ below, is calculated using the component $S_z=S_z(\sigma,\tau)$ of the bipolar Poynting vector in Eq.~(\ref{eqPoy2br}). For the right wire, the ratio can be calculated with
\begin{equation*}
f=\frac{S_z(\pi,\tau_w)}{S_z(0,\tau_w)},
\end{equation*}
and then performing the following substitutions:
\begin{enumerate}
\item The substitutions for~$h_\tau$ and~$\beta$ from Eqs.~(\ref{eqscst}) and~(\ref{eqBeta}), respectively;
\item The substitution $\tau_w=\arccosh(l/R)$;
\item The substitution $k=a/l$ from Eq.~(\ref{eqSubsk});
\item The substitution $a=\sqrt{l^{2}-R^{2}}$.
\end{enumerate}
After these steps, the simplified power density ratio is
\begin{equation*}
f=\frac{(l+R)(2l+R)}{(l-R)(2l-R)}.
\end{equation*}
When $l\gg R$, $f$ is equal to~1 and it is thus shown that the same amount of power flows locally near the inner portion of the surface and the outer portion, just as for the perfectly conducting wires in Problem~\ref{prbRSpec}.

\solprob{prbRSres2}

The ratio of power densities can be calculated using the component $S_\tau=S_\tau(\sigma,\tau)$ of the bipolar Poynting vector in Eq.~(\ref{eqPoy2br}) and the same procedure as used in the solution to Problem~\ref{prbRSres}. The ratio can also be calculated using the power density~$S_2$ given in Problem~\ref{prbS2sig}, by calculating $f=S_2(\pi)/S_2(0)$. In both cases, the result is
\begin{equation*}
f=\frac{(2l+R)}{(2l-R)}.
\end{equation*}
When $l\gg R$, $f$ is equal to~1 and it is thus shown that the same amount of power enters the wires at the inner portion of the surface and the outer portion.

\solprob{prbComplex}

Recall that a complex function $f(W)$ can be decomposed into real and imaginary parts $u(\sigma,\tau)$ and $v(\sigma,\tau)$ respectively as
\begin{equation*}
f(W)=u(\sigma,\tau)+iv(\sigma,\tau)=u+iv.
\end{equation*}
The function~$f$ is said to be complex analytic in a given region when~$u$ and~$v$ are continuously differentiable and satisfy the Cauchy-Riemann equations~\cite{SCHAUM2009} %section 3.3
\begin{equation*}
\frac{\partial u}{\partial \sigma}=\frac{\partial v}{\partial \tau},\qquad 
\frac{\partial u}{\partial \tau}=-\frac{\partial v}{\partial \sigma}.
\end{equation*}

From Eqs.~(\ref{eqVr}) and (\ref{eqvecAr0}) for the resistive wire pair, the following two expressions of the electric scalar potential~($V$) and the~$z$ component of the magnetic vector potential ($A_z$) can be obtained:
\begin{equation*}
V=\frac{\Delta V-2\rho J z}{2\arccosh(l/R)}\ln\left(\frac{\sqrt{\left(x+a\right)^2+y^2}}{\sqrt{\left(x-a\right)^2+y^2}}\right),
\end{equation*}
\begin{equation*}
A_z=\frac{\mu_0 I}{2\pi}\ln\left(\frac{\sqrt{\left(x+l\right)^2+y^2}}{\sqrt{\left(x-l\right)^2+y^2}}\right).
\end{equation*}

The potentials above can be expressed using the complex variable~$Z$ as follows:
\begin{equation*}
V=\frac{\Delta V-2\rho J z}{2\arccosh(l/R)}\ln\left(\frac{|Z+a|}{|Z-a|}\right),
\end{equation*}
\begin{equation*}
A_z=\frac{\mu_0 I}{2\pi}\ln\left(\frac{|Z+l|}{|Z-l|}\right).
\end{equation*}
Defining the complex potentials as
\begin{equation*}
\label{eqVc}
V_c=\frac{\Delta V-2\rho J z}{2\arccosh(l/R)}\ln\left(\frac{Z+a}{Z-a}\right),
\end{equation*}
\begin{equation*}
\label{eqAzc}
A_{zc}=\frac{\mu_0 I}{2\pi}\ln\left(\frac{Z+l}{Z-l}\right),
\end{equation*}
it is therefore the case that 
\begin{equation*}
V=\Re[V_c];\qquad A_z=\Re[A_{zc}].
\end{equation*}
Using the identity
\begin{equation*}
\arctanh Z=\frac{1}{2}\ln\left(\frac{1+Z}{1-Z}\right),
\end{equation*}
omitting the $i2\pi n$ additive complex constant~\cite{SCHAUM2009} %p.45
since the real part will be taken further below, the complex potentials can be re-written as
\begin{equation*}
\label{eqVc2}
V_c=\frac{\Delta V-2\rho J z}{\arccosh(l/R)}\arctanh\!\left(\frac{a}{Z}\right),
\end{equation*}
\begin{equation*}
\label{eqAzc2}
A_{zc}=\frac{\mu_0 I}{\pi}\arctanh\!\left(\frac{l}{Z}\right).
\end{equation*}
After application of the complex bipolar transformation $Z=T(W)$ given in the problem's statement, the complex potentials become
\begin{equation*}
\label{eqVc3}
V_c=-i\frac{(\Delta V-2\rho J z)W}{2\arccosh(l/R)}.
\end{equation*}
\begin{equation*}
A_{zc}=-i\frac{\mu_0 I}{\pi}\arctan\!\left(\frac{\tan\left(\frac{W}{2}\right)}{k}\right),
\end{equation*}
where the change of variables
\begin{equation*}
k=\frac{a}{l}
\end{equation*}
is used in~$A_{zc}$ to simplify further calculations. 

The complex form of the electric potential~$V_c$ is no longer required, leading to
\begin{equation*}
V=\Re[V_c]=\frac{(\Delta V-2\rho J z)\tau}{2\arccosh(l/R)},
\end{equation*}
since $\Re[-iW]=\tau$. The electric field~$\vect{E}$ can be calculated as before:
\begin{equation*}
\label{eqEbrac}
\begin{split}
\vect{E}=-\nabla V 	&= -\frac{1}{h_\tau}\frac{\partial V}{\partial\tau}\hat{\vects{\tau}} -\frac{\partial V}{\partial z}\hat{\vect{z}}\\
										&=-\frac{\Delta V-2\rho J z}{2h_\tau\arccosh(l/R)}\,\hat{\vects{\tau}} + \frac{\rho J\tau}{\arccosh(l/R)}\,\hat{\vect{z}}.
\end{split}
\end{equation*}

The complex potential~$A_{zc}$, when decomposed in real and imaginary parts $u(\sigma,\tau)$ and $v(\sigma,\tau)$ respectively, can be shown to be complex analytic in the region outside of the wires.
For analytic functions, the complex derivative is equal to~\cite{SCHAUM2009}
\begin{equation*}
\frac{d A_{zc}}{dW}=\frac{\partial u}{\partial \sigma}-i\frac{\partial u}{\partial \tau}.
\end{equation*}
The $z$ component of the magnetic vector potential is precisely the real component of $A_{zc}$,
\begin{equation*}
A_z=\Re[A_{zc}] = u(\sigma,\tau).
\end{equation*}
As such, the differentials required in the calculation of the magnetic field~$\vect{B}$ (using the curl operator) can be related to~$u$ as follows:
\begin{equation*}
\vect{B}=\vects{\nabla}\times\vect{A} = \frac{1}{h_\tau}\frac{\partial A_z}{\partial\tau}\hat{\vects{\sigma}} -\frac{1}{h_\sigma}\frac{\partial A_z}{\partial\sigma}\hat{\vects{\tau}}
=\frac{1}{h_\tau}\frac{\partial u}{\partial\tau}\hat{\vects{\sigma}} -\frac{1}{h_\sigma}\frac{\partial u}{\partial\sigma}\hat{\vects{\tau}}.
\end{equation*}
From the complex derivative further above:
\begin{equation*}
\frac{\partial u}{\partial\tau}=-\Im\left[\frac{d A_{zc}}{d W}\right],\qquad
\frac{\partial u}{\partial\sigma}=\Re\left[\frac{d A_{zc}}{d W}\right],
\end{equation*}
such that
\begin{equation*}
\vect{B}=-\frac{1}{h_\tau}\Im\left[\frac{d A_{zc}}{d W}\right]\hat{\vects{\sigma}}-
\frac{1}{h_\sigma}\Re\left[\frac{d A_{zc}}{d W}\right]\hat{\vects{\tau}}.
\end{equation*}
The required derivative is
\begin{equation*}
\frac{d A_{zc}}{d W}=-i\frac{\mu_0 I k}{(1+k^2+(k^2-1)\cos W)\pi}.
\end{equation*}
With the properties
\begin{equation*}
\Im[iZ] = \Re[Z],\qquad \Re[iZ]=-\Im[Z],
\end{equation*}
the magnetic field is thus equal to
\begin{equation*}
\label{eqBbrac}
\vect{B}=\frac{\mu_0 I}{\pi}\left(\frac{1}{h_\tau} \Re[\beta] \,\hat{\vects{\sigma}} -\frac{1}{h_\sigma} \Im[\beta] \,\hat{\vects{\tau}}\right),
\end{equation*}
where
\begin{equation*}
\label{eqbeta}
\beta=\beta(W)=\frac{k}{1+k^2+(k^2-1)\cos W}.
\end{equation*}
With the $\vect{E}$ and $\vect{B}$ fields calculated above, the Poynting vector in the space surrounding the resistive wires is
\begin{equation*}
\vect{S}=
\frac{1}{\mu_0}\left(\vect{E}\times\vect{B}\right)=
-\frac{E_z B_\tau}{\mu_0}\hat{\vects{\sigma}}
+\frac{E_z B_\sigma}{\mu_0}\hat{\vects{\tau}}
-\frac{E_\tau B_\sigma}{\mu_0}\hat{\vect{z}},
\end{equation*}
which is equal to
\begin{equation*}
\begin{split}
\vect{S} = \frac{I}{\pi\arccosh(l/R)}\bigg(\frac{\rho J \tau}{h_\sigma}\,\Im[\beta] \,\hat{\vects{\sigma}}\,&+\frac{\rho J \tau}{h_\tau}\,\Re[\beta] \,\hat{\vects{\tau}}\\
&+\frac{\Delta V-2\rho J z}{2 (h_\tau)^2}\,\Re[\beta] \,\hat{\vect{z}}\bigg).
\end{split}
\end{equation*}

The power $P_z$ flowing parallel to the resistive wires, in the space surrounding the wires, can be calculated with
\begin{equation*}
P_z=\int{S_z\,dA}=\int_{-\tau_w}^{\tau_w}\int_{0}^{2\pi}
S_z\, h_\sigma h_\tau \, d\sigma d\tau.
\end{equation*}
Since~$h_\sigma=h_\tau$, both scale factors cancel out with the ones in~$S_z$, leaving
\begin{equation*}
P_z=\frac{(\Delta V-2\rho J z)I}{2\pi\arccosh(l/R)}
\int_{-\tau_w}^{\tau_w}\int_{0}^{2\pi}\Re[\beta(\sigma+i\tau)]\, d\sigma d\tau.
\end{equation*}
The inner integral is 
\begin{equation*}
\int_{0}^{2\pi}\Re[\beta(\sigma+i\tau)]\, d\sigma = \Re\left[\int_{0}^{2\pi}\beta(\sigma+i\tau)\, d\sigma\right].
\end{equation*}
Since~$\beta$ can also be shown to be complex analytic in the region outside of the wires, the definite integral is path independent and is the same as in ordinary calculus~\cite{SCHAUM2009}. %p.117
Omitting the integration constant, the primitive~$F$ of~$\beta$ with respect to~$\sigma$ is
\begin{equation*}
\label{eqIntar1eac}
F(\sigma) =\arctan\!\left(\frac{{\tan\left(\frac{\sigma}{2}+i\frac{\tau}{2}\right)}}{k}\right),
\end{equation*}
and the definite integral is now equal to
\begin{equation*}
\label{eqIntar1fac}
F(2\pi)-F(0).
\end{equation*}
At $\sigma=0$, the primitive evaluates to
\begin{equation*}
\label{eqIntar1g0}
F(0) =\arctan\!\left(\frac{{\tan\left(i\frac{\tau}{2}\right)}}{k}\right).
\end{equation*}
At $\sigma=2\pi$, the $\tan()$ function's argument in $F(\sigma)$ is in its second branch (with a real component of~$\pi$). The result of the $\arctan()$ must therefore be offset by $+\pi$, as shown below
\begin{equation*}
\label{eqIntar1g2p}
F(2\pi)=\arctan\!\left(\frac{{\tan\left(\pi+i\frac{\tau}{2}\right)}}{k}\right)
       =\arctan\!\left(\frac{{\tan\left(i\frac{\tau}{2}\right)}}{k}\right)+\pi.
\end{equation*}
Since the first term in $F(2\pi)$ is equal to $F(0)$, it is shown that
\begin{equation*}
\Re\left[\int_{0}^{2\pi}\beta(\sigma+i\tau)\, d\sigma\right]=\pi.
\end{equation*}

The outer integral in the $P_z$ calculation is now evaluated:
\begin{equation*}
P_z=\int\limits_{-\arccosh(l/R)}^{\arccosh(l/R)}\frac{(\Delta V-2\rho J z)I}{2\arccosh(l/R)}\, d\tau 
= (\Delta V-2\rho J z)I.
\end{equation*}
Using the usual definition of the current density~$J$, the power flowing through a transverse plane at position~$z$, outside of the wires, is therefore
\begin{equation*}
P_z=I\,\Delta V-2\frac{\rho z}{\pi R^2}I^2,
\end{equation*}
which is identical to the result obtained in Eq.~(\ref{eqPzr}). The power entering the wires can also be calculated using a very similar approach.

\end{document}